\documentclass[11pt,a4paper]{article}
\usepackage{jcappub}
\usepackage{epsfig}
\usepackage{latexsym}
\usepackage{natbib}
\usepackage{url}
\usepackage{dcolumn}
\usepackage{color}
\usepackage{amsfonts,amssymb,amsmath}
\usepackage{graphicx,epsfig}
\usepackage{subfigure}
\usepackage{float}
\usepackage{mathrsfs}
\usepackage[dvipsnames]{xcolor}
\usepackage{comment}

\title{
Constraints on the mass and self-coupling of Ultra-Light Scalar Field Dark Matter using observational limits on galactic central mass}

\author[a]{Sayan Chakrabarti,}
\author[b]{Bihag Dave,}
\author[c]{Koushik Dutta,}
\author[d]{and Gaurav Goswami}

\affiliation[a]{Department of Physics, Indian Institute of Technology Guwahati,
Guwahati - 781039, India}
\affiliation[b]{School of Engineering and Applied Science, Ahmedabad University,
Ahmedabad - 380009, India}
\affiliation[c]{Department of Physical Sciences, Indian Institute of Science Education And Research Kolkata, 
Mohanpur, Nadia - 741246, India}
\affiliation[d]{Division of Mathematical and Physical Sciences, School of Arts and Sciences, Ahmedabad University,
Ahmedabad - 380009, India}

\emailAdd{sayan.chakrabarti@iitg.ac.in}
\emailAdd{bihag.d@ahduni.edu.in}
\emailAdd{koushik@iiserkol.ac.in}
\emailAdd{gaurav.goswami@ahduni.edu.in}

\abstract{
It is well known that Ultra-Light Dark Matter (ULDM), usually scalar fields of  mass $m \sim 10^{-22}~{\rm eV}$, can solve some of the outstanding problems of the Cold Dark Matter (CDM) paradigm. Such a scalar field could have non-negligible self-coupling $\lambda$.
In this work, 
using the known observational upper limit on the amount of centrally concentrated dark matter in a galaxy, we arrive at the observational constraints in the $\lambda - m$ (self coupling $-$ mass) parameter space. 
It is found that the observational limit on the mass $m$ of the ULDM depends upon the sign and strength of the self-interactions.
We demonstrate that, for $m \sim 10^{-22}~{\rm eV}$, self-coupling values of ${\cal O}(10^{-96})$ (corresponding to a scattering length of $a_s \sim 10^{-82}~{\rm m}$) can be probed using limits on the dark matter mass within 10 pc of the centre of M87 galaxy.
Our analysis suggests that if Ultra Light Axion particles (ULAs) form all of dark matter, dark matter particle mass must be less than
$\sim 6 \times 10^{-23}$ eV.
}


\arxivnumber{2202.11081}

\begin{document}
\maketitle
\flushbottom

\section{Introduction}
\label{sec:intro}

There are several reasons to believe that the Universe has a clustering, non-relativistic component which is non-baryonic in origin \cite{Planck:2018vyg}. Uncovering the fundamental nature of this component, known as dark matter (DM), is one of the major challenges in cosmology \cite{Profumo:2019ujg,Bertone:2016nfn}. 
DM could be a collisionless system of non-relativistic Bosons or Fermions, called Cold Dark Matter - a picture of DM which works well for observations at length scales greater than ${\cal O}(1) ~ {\rm Mpc}$. But at shorter distances, there exist certain observations which might be interpreted to be pointing to deviations from this picture \cite{Bullock:2017xww}. As an example, consider the core-cusp problem \cite{Flores:1994gz,Moore:1994yx}: with cold, collisionless DM, the centres of dark matter-only galactic halos \cite{Zavala:2019gpq} are expected to be more dense and more cuspy as opposed to what observations suggest \cite{Bullock:2017xww}. One way to solve such small scale problems is to work with alternative models of DM such as those in which it is described by a classical scalar field (see however, \cite{Boldrini:2021aqk}).

If the cosmological DM consists of sufficiently light spinless (i.e. Bosonic) particles \cite{Hu:2000ke}, then, owing to large occupation numbers, one expects that it can be described by a classical field \cite{Hui:2016ltb,Urena-Lopez:2019kud,Niemeyer:2019aqm,Ferreira:2020fam,Visinelli:2021uve,Hui:2021tkt}. 
Then, in the most general case, the gravitational dynamics of DM is captured by Gross-Pitaevskii-Poisson equations which result from the weak field, slowly varying, non-relativistic limit of a classical field theory consisting of Einstein gravity and a self-interacting canonical scalar field. 
In such models, it is extremely important to be able to find out observational constraints on the mass ($m$) and self-coupling ($\lambda$) of the scalar field. 

The values of mass ($m$) and self-coupling ($\lambda$) of the scalar field dark matter get constrained by many considerations.
E.g. the requirement of having the de-Broglie wavelength comparable to the size of the galaxy tells us that the mass $m$ must be roughly of the order of $10^{-22} ~{\rm eV}$ \cite{Hu:2000ke}. 
\footnote{It is worth noting that, for self-interacting dark matter particles in contrast to wavy dark matter, the constraint on self-coupling can be found in the following way: the requirement that, owing to bullet cluster observations \cite{Markevitch:2003at}, the mean free-path of the DM particles must be larger than the size of a galaxy cluster (i.e. ${\cal O}(1)~{\rm Mpc}$), we know that $\sigma / m \lesssim {\cal O}(1)~{\rm cm}^2/{\rm g}$, where, $\sigma$ is the self-interaction cross-section of DM particles. For $\lambda \phi^4$ theory, assuming $m \sim 10^{-22}~{\rm eV}$, this implies that the self coupling $\lambda < 10^{-44}$. In this context, see also \cite{Fan:2016rda}.}
Furthermore, for ultra-light axions \cite{Marsh:2015xka} with $m \sim 10^{-22}~ {\rm eV}$ and a decay constant not too far from Planck scale e.g. $f \sim 10^{17} ~{\rm GeV}$, the self coupling $\lambda$ is expected to be negative and of ${\cal O}(10^{-96})$. At this stage, we remind the reader that for attractive scalar self-interactions, the self-coupling $\lambda$ is negative while for repulsive ones, it is positive.  

While the subject of gravitational instability of self-interacting scalar fields has been studied for a very long time (see e.g. \cite{Khlopov:1985jw}), some recent studies which impose observational constraints on $m$ and $\lambda$ include \cite{Cembranos:2018ulm} (which uses Cosmic Microwave Background anisotropies and large-scale structure data to obtain $m>10^{-24}\ \text{eV}$ and $\lambda < 10^{-99}$), 
\cite{Suarez:2016eez,Chavanis:2020rdo} (which argues for the existence of ``minimum DM halo" and uses this to constrain both the mass $m$, $2.19\times 10^{-22}\ \text{eV} <  m < 2.92\times 10^{-22}\  \text{eV}$, and the scattering length, $a_s = \frac{\hbar}{mc}\frac{\lambda}{32\pi}$, for attractive interactions, they obtain $- \lambda \leq 10^{-90}$, in this context, see also, \cite{Chavanis:2019bnu,Chavanis:2019amr}),
\cite{Desjacques:2017fmf} (which uses structure formation),
\cite{Urena-Lopez:2019xri} (which uses nucleosynthesis constraints on non-thermal relativistic degrees of freedom to constraint light DM), \cite{Delgado:2022vnt} (which uses observed velocity rotation curves to obtain $\lambda \sim 10^{-90}$ for $m \sim 10^{-22}~ {\rm eV}$),
\cite{Dev:2016hxv} (which uses the change in the speed of Gravitational Waves as they pass through DM halos), \cite{Davoudiasl:2019nlo} (which uses black hole superradiance based considerations to rule out DM masses of the order of $~10^{-21}$ eV) etc.
Similarly, recently, for axionic dark matter, the axion decay constant, $f$, has been constrained using compact binary stars systems and the Earth-sun system ~\cite{KumarPoddar:2019jxe, Poddar:2021sbc}.

In addition, recently there have been studies \cite{Bar:2019pnz},  \cite{Davies:2019wgi} relating measurements of the dynamical environment of supermassive black holes (SMBH) at galactic centres and ultralight dark matter.
In \cite{Davies:2019wgi}, the authors studied a novel method to impose observational constraints on the mass of fuzzy dark matter which works at small ($\sim pc$) scales. This method was based on the following considerations:
\begin{itemize}
 \item The core of the DM halo of a galaxy is mathematically modelled by spatially localised, spherically symmetric, stationary solitonic solutions of the Schr\"{o}dinger-Poisson equations.
 \item The existence of an SMBH at the centre of the galaxy can cause the scalar field to accrete into the black hole over a time scale proportional to $m^{-6}$, where, $m$ is the mass of the scalar field; thus, for too large masses, the accretion time will be too small, providing an upper limit on the allowed values of $m$.
 \item In addition, for some galaxies, the upper limit on the amount of DM enclosed within a spherical region of some fixed radius $r_*$ from the centre is observationally known. Certain values of $m$ will produce core profiles which are consistent with these observational results while others will not be. This can provide another upper limit on the allowed values of $m$ which could be lower than the one obtained from accretion time considerations.
\end{itemize}

Since the size and mass of the dark matter core are inversely proportional to $m$, the amount of dark matter mass contained within central regions of the galaxy, which is observationally constrained, will depend on $m$ - thus, observations will constrain $m$. If the scalar field dark matter has self-interactions, for repulsive self-interactions, the core is bigger, while for attractive self-interactions, the core is smaller. This implies that the constraints which the observations of the total mass within central regions of galaxies impose on $m$ will depend on the self-coupling $\lambda$. 
Thus, it is useful to ask whether the approach presented in \cite{Davies:2019wgi} can be extended to impose observational constraints when the scalar field describing DM has non-negligible scalar self-interactions. In other words, can we find the region of $\lambda - m$ plane which gets observationally excluded by such an analysis? 

A DM candidate of enormous interest is the ultra-light axion whose mass and coupling are related in a very particular way if relic abundance constraints are to be satisfied \cite{Marsh:2015xka}. It is thus important to ask whether the method we present in this paper could be used to constrain the parameter space of ultra-light axion dark matter.

This paper is organised as follows:
in section \ref{sec:reminder}, we briefly remind the reader how Gross-Pitaevskii-Poisson equations can be obtained as the weak field, slow variation, non-relativistic limit of a classical field theory consisting of Einstein gravity and a self-interacting canonical scalar field. In addition, in this section, we introduce the solutions which act as models of cores of DM halos, identify the parameters in the problem at hand and note some basic details about the astronomical object we use to demonstrate our methods in this paper. Next, in section \ref{sec:machinery}, we develop the detailed machinery to solve Gross-Pitaevskii-Poisson equation.
Then, in section \ref{sec:constraints}, we calculate the amount of DM contained within the central region of a soliton core for arbitrary $m$ and $\lambda$  and find out which regions of $\lambda - m$ plane get excluded by the considerations similar to \cite{Davies:2019wgi}. 
Finally, in section~\ref{sec:conclusion}, we summarise and conclude with a brief discussion. In addition, in two appendices at the end of the paper, we divulge the detailed reasoning which explains the dependence of various quantities of interest on other quantities of interest, thereby clarifying some of the features of the method we present.

{\bf Notation:}
In the following, we shall often work with natural units in which $\hbar = c = 1$ but occasionally, we shall put back factors of 
$\hbar$ and $c$. Note that we denote reduced Planck mass by $M_{pl}$ i.e. $M_{pl} = \sqrt{ \hbar c / 8 \pi G} \approx 2.4 \times 10^{18} ~ {\rm GeV}$. Similarly, the symbol $M_{\odot}$ stands for solar mass i.e. $M_{\odot} \approx 2 \times 10^{33} g$.

\section{A quick reminder of basics}
\label{sec:reminder}
\subsection{Scalar field dark matter: from general relativity to Gross-Pitaevskii-Poisson equations}

In this work, we are interested in the classical field theory with action
\begin{equation}
 S = \int d^4 x \sqrt{-g} \left( \frac{M_{\rm pl}^2}{2} R - \frac{1}{2} g^{\alpha \beta} \partial_{\alpha} \varphi \partial_{\beta} \varphi - U(\varphi) \right) \; ,
\end{equation}
where, $U(\varphi) = \frac{m^2 \varphi^2}{2} + \frac{\lambda \varphi^4}{4 !}$. If cosmological DM is to be described by the non-relativistic dynamics of the scalar field $\varphi$, we wish to observationally constraint the parameters $m$ and $\lambda$.
Varying the above action w.r.t. ${\varphi}$ will give the equation of motion of the scalar field in any spacetime
\begin{equation} \label{eq:EOM_1}
    \partial_\mu \left[ \sqrt{-g}~ g^{\mu \nu}~ \partial_\nu {\varphi} \right]  = \sqrt{-g} ~U'({\varphi}) \; , 
\end{equation}
where, $U'({\varphi})$ is the derivative of $U$. 
For a regime with weak gravity, the metric takes up the form 
\begin{equation} \label{eq:metric}
 ds^2 = - \left(1+ {2 \Phi} \right) (dx^0)^2 + \left(1- {2 \Phi} \right) \delta_{ij} dx^i dx^j \; ,
\end{equation}
where, ${\Phi} \ll 1$ and has only spatial variation. 
Using equation (\ref{eq:EOM_1}) and (\ref{eq:metric}), one finds that
\begin{equation}
    \partial_0^2 {\varphi} - \nabla^2 {\varphi} + U'({\varphi}) = 
    2 \Phi \left[ 2 \partial_0^2 {\varphi} + U'({\varphi}) \right] \; .
\end{equation}
Similarly, we are interested in the situations in which the scalar field dynamics can be adequately described by non-relativistic equations i.e., the scalar field has slow spatial and temporal variation. 
In the absence of gravity and scalar self-interactions, each Fourier mode of the scalar field oscillates with frequency 
${\omega}_{\bf k} = m \left(1 + {\bf k}^2 / m^2 \right)^{1/2}$ and slow variation means that the oscillation frequency is dominated by $\omega_k = m$ save for small corrections.
This suggests that, in order to successfully take the non-relativistic limit, we introduce a complex scalar field $\Psi(t,\vec{x})$, defined by
\begin{equation}
 \varphi (t, {\vec x}) = \frac{1}{\sqrt{2} m} \left[ e^{-im t} \Psi (t, {\vec x}) + {\rm c.c.} \right] \; ,
\end{equation} \label{eq:Psi}
here, the new field $\Psi$ captures the dynamics of $\varphi$ over and above the time evolution captured by $e^{-im t}$
and slow variation means that (a) the following hierarchy is maintained:
\begin{equation}
 \Psi ~\gg~ m^{-1} ~ {\dot \Psi} ~\gg~ m^{-2} ~ {\ddot \Psi} \gg \cdots \; ,
\end{equation}
and, (b) when we are interested in time scales large as compared to the Compton time $m^{-1}$, any term which is highly oscillatory (e.g. $e^{-2imt}$ etc) will average out to zero. 
Note that equation (\ref{eq:Psi}) does not define $\Psi$ uniquely and its phase can be freely chosen.

Under this weak gravity and slow variation approximation, equation of motion of the field $\Psi$ and Einstein equations will yield the Gross-Pitaevskii-Poisson equations, \cite{Pitaevskii:2016book, Chavanis:2011mrr}

\begin{eqnarray} 
     i \frac{\partial \Psi}{\partial t} &=&  -\frac{\nabla^2}{2m} \Psi + m \Phi  \Psi +  \frac{\lambda}{8 m^3} |\Psi|^2 \Psi ~+~ \cdots \label{eq:GrossPitaevskii} \\
    \nabla^2 \Phi &=& \frac{|\Psi|^2}{ 2 M_{\rm pl}^2 } ~+~ \cdots \; . \label{eq:Poisson}
\end{eqnarray}

In the absence of self-gravity, we do not have eq.~(\ref{eq:Poisson}), the $m \Phi  \Psi$ term in eq.~(\ref{eq:GrossPitaevskii}) will be absent and we recover Gross-Pitaevskii equation (i.e. non-linear Schr\"{o}dinger equation or the limit of Ginzburg-Landau equations in the absence of electromagnetic field). Alternatively, in the absence of the self-interactions, this system reduces to what is usually referred to as the Schr\"{o}dinger-Poisson system. The ``$\cdots$" in both these equations stand for ``higher-order corrections" (see e.g. \cite{Salehian:2021khb} for a recent discussion) which we completely ignore for now. We shall have more to say about them in a later section.

\subsection{Soliton solutions of Gross-Pitaevskii-Poisson equations as cores of DM halos}

One solution of the core-cusp problem \cite{Flores:1994gz,Moore:1994yx,Bullock:2017xww} is that the dark matter is ultra-light. 
If the mass of DM particles is sufficiently low, the corresponding de-Broglie wavelength is sufficiently large and hence, well inside the galaxies, they form a Bose-Einstein Condensate - a model of DM known as Fuzzy Dark Matter (FDM) \cite{Hu:2000ke}. It is these cores of fuzzy Dark Matter halos that we shall model using non-relativistic scalar field.

\subsubsection{Soliton solutions}

We shall be interested in solutions of the system of equations (\ref{eq:GrossPitaevskii}) and (\ref{eq:Poisson}) which are (a) stationary in the sense that not only is $\Phi$ time-independent (an assumption we already made in the previous subsection), the field $\Psi$ is also of the form $\Psi(t,\vec{x}) = \phi(\vec{x}) e^{\frac{-i \gamma t}{\hbar}}$ 
with $\phi$ real, 
(b) spherically symmetric, so that $\phi(\vec{x}) = \phi(r)$, (c) spatially localised, (d) nodeless (so that the solution has minimum gradient energy), and, 
(e) regular everywhere. 
We shall call such solutions ``solitons" and they will serve as models for cores of the galactic DM halos. 

Moreover, we shall consider two cases - one in the absence of a black hole at the centre of the soliton and the other with a black hole \cite{Davies:2019wgi}. The presence of a black hole will be taken care of by replacing the $m \Phi  \Psi$ term in Gross-Pitaevskii equation (\ref{eq:GrossPitaevskii}) by $m \left( \Phi - \frac{G M_\bullet}{r} \right) \Psi$, where, $M_\bullet$ is the mass of the black hole. 
Finally, we shall consider scalar self-interactions of different strengths and signs. 

\subsubsection{Parameters}

In eqs.~(\ref{eq:GrossPitaevskii}) and (\ref{eq:Poisson}), the only free (i.e. adjustable) parameters of the underlying fundamental theory are $m$ and $\lambda$, which we wish to constrain. In addition, in the presence of a central black hole, the mass of the black hole $M_\bullet$ also acts as a relevant parameter.
Since the solutions we study will serve as models of cores of DM halos, the object specific parameters which are important are:
the mass of the DM halo ($M_{\rm halo}$), the mass of its core $M$. Since the core of the DM halo will be modelled by a ``soliton" solution (as defined in the last paragraph), $M$ will also be sometimes called the mass of the soliton.

\subsubsection{M87 - an example case to illustrate the method}
\label{sec:m87}

In order to demonstrate the method that we study in the present work, we use the example of M87 (also known as NGC4486), a supergiant elliptical galaxy in the Virgo cluster. It is known to contain a SMBH of mass $ M_{\bullet} = 6.5 (\pm 0.7) \times 10^{9} M_{\odot}$ at its centre. The mass of this SMBH was recently determined by its imaging by the Event Horizon Telescope \cite{EventHorizonTelescope:2019dse}. Following \cite{Hui:2016ltb} and \cite{Davies:2019wgi}, we shall assume that the halo mass of M87 galaxy is $M_{\rm halo} = 2 \times 10^{14} M_{\odot}$. Similarly, we shall assume that the region within $10 ~{\rm pc}$ of the centre of M87 galactic halo contains a soliton mass not more than $10^{9} M_{\odot}$.

\section{Modelling realistic cores}
\label{sec:machinery}

\subsection{Dimensionless variables}
\label{sec:dimless}
In order to proceed, we can introduce dimensionless variables: for the distances i.e. 
$ {\hat r} \equiv \frac{mc r}{\hbar} $ (i.e. distance is measured in units of reduced Compton wavelength), for the gravitational potential, ${\hat \Phi} = \Phi/c^2$, for the ``wave function" ${\hat \phi} = \frac{ \sqrt{4 \pi G} \hbar }{m c^2} \phi $, for the ``energy" ${\hat \gamma} = \gamma / mc^2$ and 
a measure of the ``strength" of the black hole 
\begin{equation} \label{eq:alpha}
{\hat \alpha} = \frac{GM_\bullet m}{\hbar c} = 4.87 \times 10^{-4} ~ \left(\frac{M_\bullet}{6.5 \times 10^9 ~ M_{\odot}} \right) \left(\frac{m}{10^{-22} ~ {\rm eV} } \right) \; ,
\end{equation} 
see e.g. \cite{Davies:2019wgi}. 
For the self-coupling, it is useful to define the following variable
\begin{equation}\label{eq:lambdac} 
{\hat \lambda} =  \frac{\lambda}{8} \left(\frac{m}{M_{pl}}\right)^{-2} = \left( \frac{ \lambda}{1.35 \times 10^{-98}} \right) \left( \frac{m}{10^{-22}~ {\rm eV} } \right)^{-2} \; ,
\end{equation}
where, $M_{pl}$ is the reduced Planck mass.
It is worth noting that for ${\hat \lambda} \sim {\cal O}(1)$ and for $m \sim {\cal O}(10^{-22})~{\rm eV}$, the value of self-coupling which can be probed is incredibly small i.e. $\lambda \sim 10^{-98}$.
In terms of these new variables, Gross-Pitaevskii-Poisson equations become
\begin{eqnarray} 
 \frac{1}{2} {\hat \nabla}^2 {\hat \phi} &=& {\hat \Phi}{\hat \phi} - {\hat \gamma} {\hat \phi} - \frac{\hat \alpha}{\hat r} {\hat \phi} + 2 {\hat \lambda} {\hat \phi}^3 + \cdots  \; , \label{eq:GP_dimless}\\
  {\hat \nabla}^2 {\hat \Phi} &=&{\hat \phi}^2 + \cdots \; , \label{eq:P_dimless}
\end{eqnarray}
and we shall impose the boundary conditions ${\hat \phi}({\hat r}=\infty) = 0,~ {\hat \phi} ' ({\hat r} = 0) = 0, {\hat \Phi}({\hat r} = 0 ) = 0,~ {\hat \Phi} ' ({\hat r} = 0) = 0$.
At this stage, the parameters $\hat \alpha$ and ${\hat \lambda}$ are free and their values are to be chosen before we can proceed to solve the system. This is not true about $\hat \gamma$ $-$ for a chosen $\hat \alpha$ and ${\hat \lambda}$, the value of the parameter $\hat \gamma$ is determined by the boundary conditions and the requirement of having no nodes in the desired solution. 

\begin{figure}
  \includegraphics[width = 0.92\textwidth]{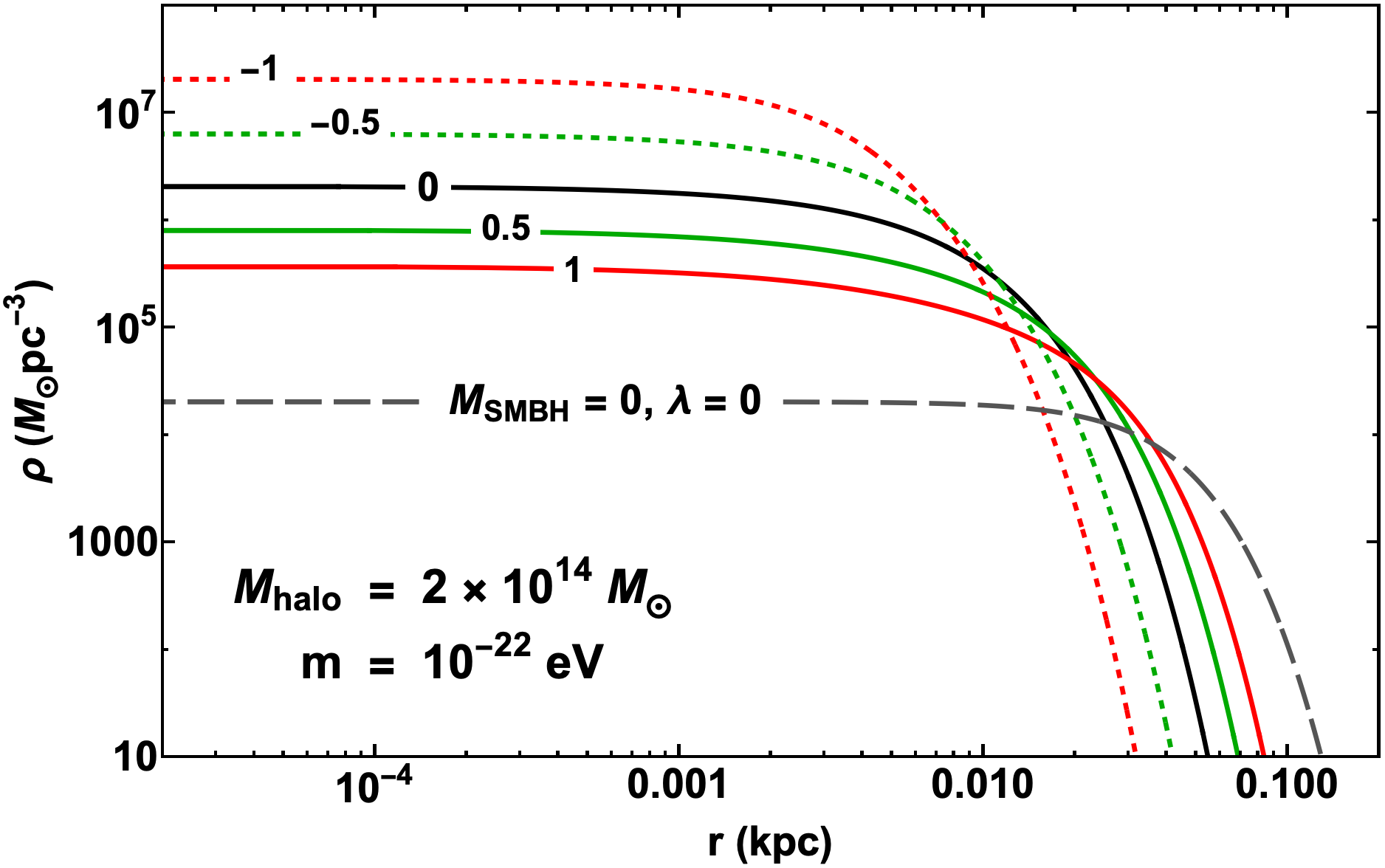}
  \caption{
Density profiles for the core of M87 halo for fixed value of Dark Matter mass, $m$, and various values of self-coupling $\lambda$. The black dashed curve at the bottom corresponds to a halo without a SMBH at the centre while assuming no self-interactions of DM. The other curves show the density profile but in the presence of a SMBH at the centre. The label on each curve is the value of $\hat{\lambda}_{ini}$ (defined in appendix~\ref{sec:parameter}). 
The various ${\hat \lambda}_{ini}$ values correspond to the following  $\lambda$ values (top to bottom): $-6.2 \times 10^{-95} $, $-5.6 \times 10^{-95}$, $0$, $1.6 \times 10^{-94}$, $4.6 \times 10^{-94}$. 
It can be seen that repulsive self-interactions expand the soliton while attractive self-interactions compress it.} 
  \label{fig:densityprofile}
\end{figure}

Thus, once $\hat \alpha$ and ${\hat \lambda}$ are specified, the system of equations can be readily solved by using the familiar shooting method, see e.g. \cite{Giordono:2006} and \cite{Davies:2019wgi}. The density profiles obtained from such solutions are shown in figure \ref{fig:densityprofile}. 
In section~\ref{sec:scaling}, we discuss which values of $\hat \alpha$ and ${\hat \lambda}$ need to be used in the solution of eqs. (\ref{eq:GP_dimless}) and (\ref{eq:P_dimless}) in order to obtain realistic solitonic solutions such as those shown in figure \ref{fig:densityprofile}.

Given any solution of the system, one can calculate the mass of a spherically symmetric soliton i.e. $M$
using
\begin{equation} \label{eq:mhat}
 {\hat M} = \frac{GMm}{\hbar c} = \int_0^{\infty} d{\hat r} ~ {\hat r}^2 {\hat \phi}^2 + \cdots  \; ,
\end{equation}
is the dimensionless mass of the soliton. As we noted earlier, the contribution denoted by ``$ \cdots$" comes from higher order corrections which we shall mostly ignore except in the next section.

\subsection{The solutions with ${\hat \phi} ({\hat r} = 0) \sim 1$}
\label{sec:theoretical}

It is easiest to find the solutions of the dimensionless Gross-Pitaevskii-Poisson equations (\ref{eq:GP_dimless}) and (\ref{eq:P_dimless}) when we impose the additional requirement that ${\hat \phi} ({\hat r} = 0) \sim 1$. Note that for numerical work, we shall always work with ${\hat \phi} ({\hat r} = 0) = 1$ but the arguments below apply to any ${\cal O}(1)$ value of ${\hat \phi} ({\hat r} = 0)$. Such solutions have many nice properties relevant for numerical work since all the relevant quantities are ${\cal O}(1)$ but, as we note below, they are highly unrealistic for the purpose of modelling real solitons. We shall call these solutions ``theoretical solitons".

\noindent The definition ${\hat \phi} = \frac{ \sqrt{4 \pi G} \hbar }{m c^2} \phi $ implies that the central density of the soliton will be 
\begin{equation}
 \rho = 2 \left( \frac{M_{pl}}{m} \right)^2 \frac{m}{(\hbar/mc)^3} \sim 10^{14}
\frac{M_{\odot}}{(pc)^3}  \; ,
\end{equation}
where we have used only the leading contribution to density (i.e. we have ignored contributions of the higher order corrections and scalar self-interactions). This density corresponds to a very large mass (of the order of halo mass for e.g. M87, see section~\ref{sec:m87}) within a volume of just $(pc)^3$, hence
solutions found by assuming ${\hat \phi}(0) \sim 1$ have too large densities.

A similar argument can be made about the size of the soliton: 
in the absence of central BH, set ${\hat \alpha} = 0$ in the equation (\ref{eq:GP_dimless}), if the size of the system, in units of $\hbar / mc$, is $\hat L$, then, Poisson's equation tells us that ${\hat \Phi} \sim {\hat L}^2 {\hat \phi}^2$, assuming that all the terms in 
Gross-Pitaevskii equation have the same order of magnitude. This will imply that ${\hat L} \sim 2^{-1/4} {\hat \phi}^{-1/2}$.
If ${\hat \phi} \le 1$, as is the case for theoretical soliton, i.e. if ${\hat \phi}$ is ${\cal O}(1)$, then, ${\hat L}$ will also be ${\cal O}(1)$. Thus, the size of theoretical soliton, in the absence of central BH, will be of the order of ULDM / FDM Compton wavelength $\hbar / mc$.
On the other hand, in the presence of central BH: If ${\hat \alpha} \ll 1$, then, the ${\hat \Phi}{\hat \phi}$ term will still dominate over 
$\frac{\hat \alpha}{\hat r} {\hat \phi}$ term in Gross-Pitaevskii equation and the size $\hat L$ will still be ${\cal O}(1)$. 
When the $\frac{\hat \alpha}{\hat r} {\hat \phi}$ term dominates, the size will be ${\hat L} \sim 1 / {\hat \alpha}$.
But, this happens only when ${\hat \alpha}$ is at least ${\cal O}(1)$, so, the dimensionless size $\hat L$ will still remain ${\cal O}(1)$.
We know that the size of the soliton must be larger than Compton wavelength, in fact it is expected to be of the order of de-Broglie wavelength of the ULDM or FDM particles.

Using eq.~(\ref{eq:mhat}) and a similar reasoning it is possible to convince oneself that the dimensionless mass $\hat M$ will be ${\cal O}(1)$. This corresponds to a soliton mass of the order of $10^{12} ~M_{\odot}$ which is much larger than the corresponding value for e.g. M87 i.e. $M \lesssim 10^9 M_{\odot}$ (see section~\ref{sec:m87}).

Finally, since the dimensionless soliton mass $\hat M$ is ${\cal O}(1)$, the gravitational radius of the soliton turns out to be
\begin{equation}
 R_G = \frac{G M}{c^2} = {\hat M} ~ \frac{\hbar}{mc} \sim \frac{\hbar}{mc} \; .
\end{equation}
Thus, the gravitational radius for a soliton with such characteristics is comparable to its size, so, general relativistic effects can not be ignored (even in the absence of a central black hole). 
Real solitons are expected to be much bigger, much less dense and much lighter.

\subsection{Scaling transformations and trustworthy regime}

It is easy to see that for Gross-Pitaevskii-Poisson equations, eqs.~(\ref{eq:GP_dimless}) and (\ref{eq:P_dimless}), ignoring the ``$\cdots$" terms,
a scaling transformation of the form
\begin{eqnarray}
 {\hat r}  &\rightarrow&   s ~ {\hat r} \; , \\
 {\hat \phi}  &\rightarrow& \frac{1}{s^{2}} ~ {\hat \phi} \; , ~{\hat \Phi} \rightarrow \frac{1}{s^{2}}~ {\hat \Phi} \; , ~ {\hat \gamma} \rightarrow \frac{1}{s^{2}}~ {\hat \gamma} \; , \\
 {\hat \alpha} &\rightarrow& \frac{1}{s}~ {\hat \alpha} \; , \\
 {\hat \lambda} &\rightarrow& s^2~ {\hat \lambda} \label{eq:lambdatrans} \; ,
\end{eqnarray}
causes each term to get scaled by a factor of $1/s^4$. In other words, such a transformation leaves the system of equations invariant.
If $s>1$, such a transformation renders the soliton lighter, bigger and less dense. 

Even if this transformation of ${\hat \lambda}$ is accepted, the higher order terms (denoted by ``$\cdots$" in previous expressions) in Poisson's equation e.g. ${\hat \lambda} ~ {\hat \phi}^4$, $- 3 {\hat \Phi} {\hat \phi}^2$ as well as those in Gross-Pitaevskii equation will violate the scaling symmetry as they scale by a factor of $1/s^6$ as compared to all the other terms which scale as $1/s^4$. But, this also implies that for sufficiently large $s$, these terms shall both become negligibly small as compared to the rest of the terms.
Thus, the regime with $s \gg 1$ is highly desirable especially because it is reliably trustworthy. 
Thus, in such a trustworthy regime, all the higher order terms (denoted by ``$\cdots$" in all the previous expressions which have them) can be completely ignored and we find that
\begin{equation} \label{eq:Mtrans}
 {\hat M} \rightarrow \frac{1}{s}~ {\hat M} \; ,
\end{equation}
i.e. the dimensionless soliton mass $\hat M$ scales just like the quantity ${\hat \alpha}$ defined by eq.~(\ref{eq:alpha}).

\subsection{The amount of scaling}

For theoretical solitons, having fixed the values of $\hat \alpha$ and ${\hat \lambda}$, once the solutions ${\hat \phi}({\hat r})$ and ${\hat \Phi}({\hat r})$ are obtained, one can readily find the dimensionless soliton mass, $\hat M$, using eq.~(\ref{eq:mhat}).
One thus needs to do the following:
\begin{itemize}
 \item start with some initial values ${\hat \alpha} = {\hat \alpha}_{ini}$ and ${\hat \lambda} = {\hat \lambda}_{ini}$, while solving eqs.~(\ref{eq:GP_dimless}) and (\ref{eq:P_dimless}) and obtain the theoretical solitons,
 \item use the scaling transformations to turn theoretical solitons into real solitons whose parameters take realistic values. 

\end{itemize}
In other words, the scaling transformation parameter ``$s$" needs to be chosen to ensure that $\hat M$ of the real soliton as well as the corresponding $\hat \alpha$ are equal to the observed values. The way to do this is explained in detail in 
appendices \ref{sec:scaling}
and 
\ref{sec:parameter}.
Thus, the correct value of scaling parameter $s$ will be such that 
\begin{equation} \label{eq:s}
s = \frac{{\hat M}_{ini}}{{\hat M}_{\rm emp}} = \frac{{\hat \alpha}_{ini}}{{\hat \alpha}_{\rm emp}} \; ,
\end{equation}
where, ${\hat M}_{ini}$ and ${\hat \alpha}_{ini}$ are defined in the appendix \ref{sec:parameter}.
Since the soliton size scales as $s$, the ratio of the physical size $L$ of the actual soliton and $m^{-1}$, the size of ``theoretical" soliton, will also be related by
\begin{equation} \label{eq:s2}
s = \frac{L}{m^{-1}} \; .
\end{equation}
It is also worth noting that since ${\hat M}_{ini}$ is large compared to ${\hat M}_{\rm emp}$, we must have $s \gg 1$ so that ${\hat \alpha}_{trans} \ll {\hat \alpha}_{ini}$.

\subsubsection{Typical values of interest}
\label{sec:typical}

At this stage, it is useful to compare the strength of gravity and scalar self-interactions.
Using Poisson's equation (\ref{eq:Poisson}), it is easy to see that, in Gross-Pitaevskii equation (\ref{eq:GrossPitaevskii}), the ratio of gravitational term $m \Phi \Psi$ and the self-interaction term $\frac{\lambda}{8m^3} |\Psi|^2 \Psi$ will be 
\begin{equation}
 \frac{\rm gravitation}{\rm scalar~ interaction} = \frac{4}{\lambda} \left( \frac{m}{M_{pl}} \right)^2 \left( \frac{L}{m^{-1}} \right)^2 \; ,
\end{equation}
i.e. as the size of the object increases, gravity becomes far more important than self-interactions.
This is expected since the quartic self-interactions of the scalar field are contact interactions while gravity is a long range force.
Now using eqs.~(\ref{eq:s2}) and (\ref{eq:lambda}), this ratio is also the same as
\begin{equation}
 \frac{\rm gravitation}{\rm scalar~ interaction} = \frac{1}{2 {\hat \lambda}_{ini}} \; ,
\end{equation}
so that if gravity always dominates over scalar self-interactions then we should have 
\begin{equation}
  - \frac{1}{2} \le {\hat \lambda}_{ini} \le \frac{1}{2} \; . 
\end{equation}
In this work, we will not focus on the case in which self-interactions dominate over gravity, hence, the range of ${\hat \lambda}_{ini}$ of interest to us is given by the above equation.
Needless to say, if gravity is negligibly small, we go from the domain of Gross-Pitaevskii-Poisson equations to the domain of non-linear Schr\"{o}dinger equation.
Since we are going to be interested in Ultra Light Dark Matter, we work with the following range of $m$
\begin{equation}
10^{-24}~{\rm eV} \le m \le 10^{-21}~{\rm eV} \; . 
\end{equation}
As stated, for the purpose of this work, we restrict our attention to the case of M87 galaxy, so we shall keep the parameter $M_{\rm halo}$ to the value $2 \times 10^{14} M_{\odot}$ (see section~\ref{sec:m87}).

\section{Imposing constraints in $m - \lambda$ plane}
\label{sec:constraints}

Having developed the machinery to probe the self-coupling, we now go back to the method proposed by Davies and Mocz \cite{Davies:2019wgi} to impose constraints on the mass of the fuzzy dark matter and find out what it can teach us about the joint constraints on the mass $m$ and the scalar self-coupling $\lambda$.

As discussed in the Introduction  section~\ref{sec:intro}, the first step involved in the method studied by \cite{Davies:2019wgi} is to model the soliton using stationary, spherically symmetric solutions of Schr\"{o}dinger-Poisson equations. For the case of our interest, this step involved modelling the soliton using similar solutions of Gross-Pitaevskii-Poisson equations. The machinery outlined in the previous section was developed to obtain solitonic solutions of Gross-Pitaevskii-Poisson equations for any value of $m$, ${\hat \lambda}_{ini}$ and for a galaxy with halo mass $M_{\rm halo}$.

The next step involves the realisation that the mass of the ultra light dark matter will be constrained simply because of considerations based on the accretion of the scalar field into the central black hole. Again, following \cite{Davies:2019wgi}, we shall simply remind the reader that in the presence of a SMBH at the centre of the scalar field soliton, the scalar field will accrete into the black hole. 
According to the calculations of \cite{Barranco:2011eyw} and \cite{Barranco:2017aes}, the corresponding accretion time  will be proportional to $t_{acc} \propto M_{\bullet}^{-5} m^{-6}$ and hence, for a given black hole mass there is always a value of ULDM mass, $m$ such that, for all masses larger than that, the accretion time is too small compared to cosmological time scales. This consideration alone imposes a constraint on the mass of the DM - specifically, it will provide an upper limit on $m$. 
The calculations of \cite{Barranco:2011eyw} and \cite{Barranco:2017aes} do not deal with a scalar field with self interactions, but, since this aspect is not the main point of our work, we shall demonstrate all our results by simply assuming that the accretion based constraints hold good even when the scalar field has non-negligible scalar self-interactions. 

The third step involves calculating the DM mass within some central region of a galactic halo core using the solutions we already have. Needless to say, this mass will depend on the free parameters such as $m$ and $\lambda$ and comparison with observations will then exclude some regions in the $\lambda - m$ parameter space. 

Keeping this in mind, we now turn to the calculation of the DM mass within a central region of some observable size.

\subsection{The amount of centrally concentrated dark matter for a soliton core}
\label{sec:M&D}

Using the solitonic solutions we have obtained (see figure~\ref{fig:densityprofile}), we need to calculate the DM mass, $M_{r < r_*}$, within a spherical region of some fixed radius, say $r_{*}$, from the centre of the galactic halo. Note that $M_{r < r_*}$ could be calculated irrespective of whether there is SMBH at the centre of the galaxy or not.
The dimensionless variable specifying the radius of the region will be
 \begin{equation} \label{eq:rhatstd}
 {\hat r}_* = \frac{r_*}{\hbar / mc} \; .
\end{equation} 

As we saw in section~\ref{sec:m87}, it is assumed that for M87, a soliton mass of only $10^9~M_{\odot}$ can be contained within a distance of 10 pc of the SMBH. In other words, $r_*$ is taken to be 10 pc - we shall continue to work with this number.
The discussions in the last section suggest that there must exist a scaling parameter $s$ which takes us from the theoretical soliton to the real soliton. 
When we use the empirical relations to find the value of $s$, it is found that the value of $s$ in the presence of black hole is smaller than the value of $s$ in the absence of black hole (though both of them are larger than 1) i.e. the presence of a central black hole squeezes the soliton, as was found in \cite{Davies:2019wgi} and also can be seen in figure~\ref{fig:densityprofile}.

In section~\ref{sec:theoretical}, we had argued that the unscaled size of the soliton i.e. the size of theoretical soliton is ${\cal O}(m^{-1})$. Note that the size of the soliton could be defined more precisely e.g. one could define the distance within which 95\% of the mass of the soliton exists to be the size of the soliton. 
Using any such definition of the size of the soliton, if the unscaled dimensionless size of the soliton is ${\hat r}_{\rm sol}$, then, eq.~(\ref{eq:s2}) suggests that the dimensionless size of the soliton after scaling will be $ s {\hat r}_{\rm sol}$. 

Now, there are two possibilities
\begin{itemize}
 \item $ s {\hat r}_{\rm sol} > {\hat r}_* $ i.e. only a fraction of the soliton is inside the region of interest. In this case, the mass within this region should be
\begin{equation} \label{eq:M_limited}
M_{r < r_*} = \left( \frac{\hbar c}{G m} \right) \int_0^{{\hat r}_*} d{\hat r} ~ {\hat r}^2 {\hat \phi}^2 \; .
\end{equation}
where, it is to be noted that the quantities used inside the integral are scaled.
 \item $s {\hat r}_{\rm sol} < {\hat r}_* $ i.e. the entire soliton is inside the region of interest. In this case, the mass within this region is
\begin{equation} \label{eq:M_full}
M_{r < r_*} = \left( \frac{\hbar c}{G m} \right) \int_0^{s {\hat r}_{\rm sol}} d{\hat r} ~ {\hat r}^2 {\hat \phi}^2 \; ,
\end{equation}
where, again the quantities used inside the integral are scaled. Since the entire soliton is well inside the region of interest, at $s {\hat r}_{\rm sol}$ the field takes up very small values and so if we replace the upper limit in the integral in the last expression by $\infty$, the error in evaluating it is expected to be very small. 
In this approximation, $M_{r < r_*}$ will be the same as the total mass of the soliton itself (recall the discussion just before eq.~(\ref{eq:mhat})).
\end{itemize} 
This is how the quantity $M_{r < r_*}$ can be evaluated by solving eqs.~(\ref{eq:GP_dimless}) and (\ref{eq:P_dimless}), using the machinery developed in the last section. Let us suppose that for a given galaxy, the halo mass $M_{\rm halo}$ and black hole mass $M_{\bullet}$ are known, then, in principle, we could find out $M_{r < r_*}$ for a particular choice of $m$ and $\lambda$. 

\begin{figure}[t]
\centering
\begin{subfigure}
\centering
  \includegraphics[width = 0.8\textwidth]{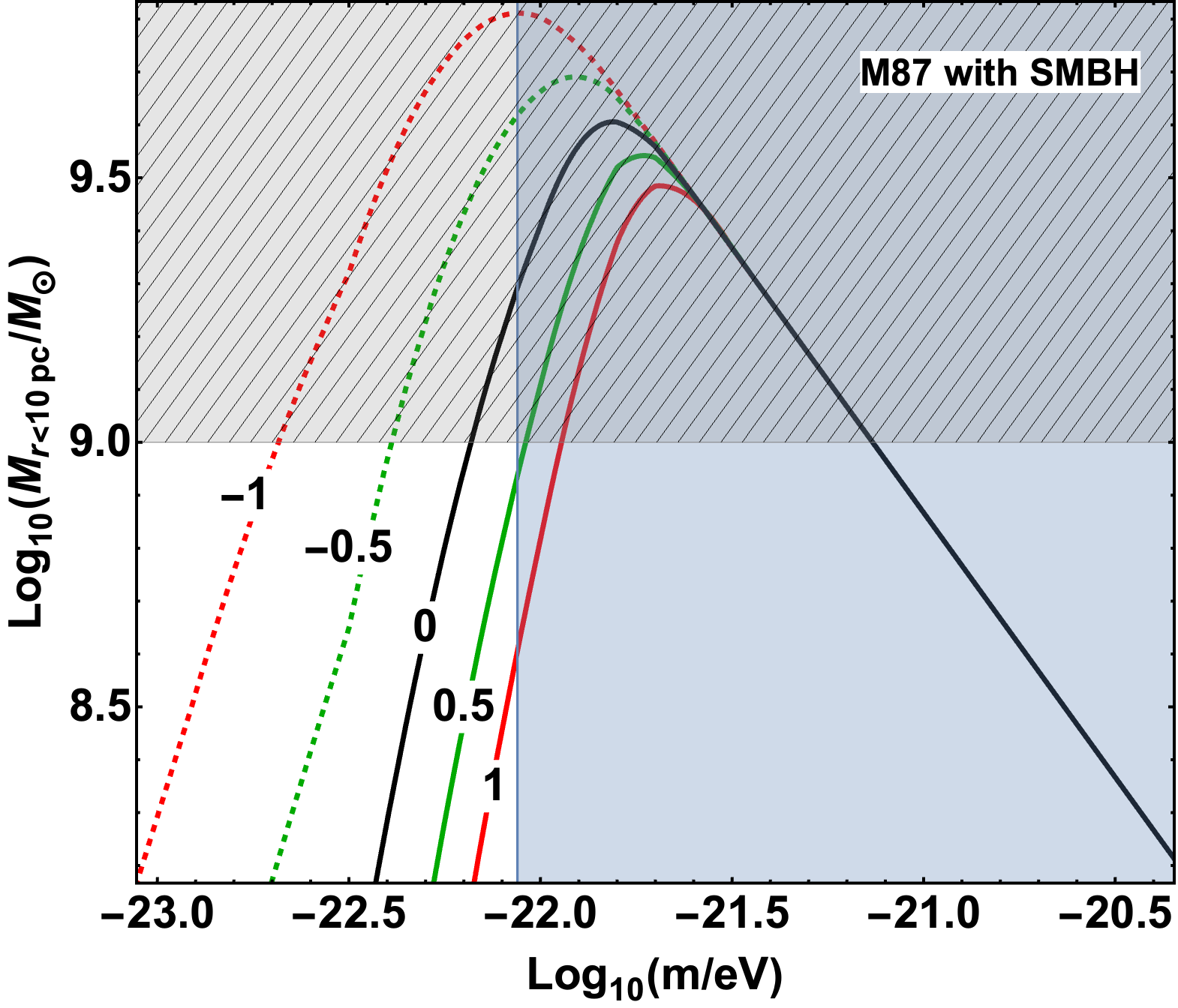}
  \caption{The curves represent the mass within a distance of 10 pc of the centre of M87 galaxy ($M_{\rm halo} = 2\times 10^{14} ~ M_{\odot}$) against values of mass of Ultra Light Dark Matter i.e. $m$ and for a few different values of ${\hat \lambda}_{ini}$ in the presence of a central SMBH at the centre of the soliton. The shaded and marked regions are excluded from observations, see text for details. In figure~\ref{M87WithBH},  the solid curve labelled ``0" corresponds to the results of \cite{Davies:2019wgi}.}
  \label{M87WithBH}
\end{subfigure}
\end{figure}

To begin with, we proceed in the following manner - we fix ${\hat \lambda}_{ini}$ and $m$, obtain the density profile (such as in figure~\ref{fig:densityprofile}) and find $M_{r < r_*}$. Then, we repeat this process for different values of $m$. It is in this manner that each curve in figure~\ref{M87WithBH} has been obtained. It is worth noting that though all points on a given curve in figure~\ref{M87WithBH} have the same ${\hat \lambda}_{ini}$, each point on a given curve in figure~\ref{M87WithBH} corresponds to a different value of $\lambda$ since the corresponding value of $m$ is different (see eq.~(\ref{eq:lambdac})).
Towards the end of appendix~\ref{sec:parameter}, we note that each curve in figure~\ref{fig:densityprofile} has different value of the scaling parameter $s$. 
In comparison, in figure~\ref{M87WithBH}, the value of the scaling parameter $s$ is not only different for different curves, it is also different for different points on the same curve, in this context, see appendix~\ref{sec:m&lambda} for more details.

Thus, for any given DM halo (i.e. fixed $M_{\rm halo}$ and $M_{\bullet}$), one can obtain a plot such as figure \ref{M87WithBH}.
We have explained the origin of the shapes of various curves in figure~\ref{M87WithBH} in appendix~\ref{sec:Shape1}. 
The only noteworthy facts at this stage are (a) for any object (such as M87) and for any choice of ${\hat \lambda}_{ini}$, one can obtain a plot of 
$M_{r < r_*}$ against $m$, (b) due to accretion time considerations and due to observational limits on mass within $r_*$ of the centre of the halo, there will be regions of these plots which are observationally disallowed,
(c) the combinations of $m$ and ${\hat \lambda}_{ini}$ which correspond to the portion of the curves that lie within either of the shaded regions will also get observationally ruled out.

\subsection{Results}

In figure~\ref{M87WithBH}, note that $m$ values in the shaded region (i.e. those to the right of the vertical line) will be excluded based on accretion time considerations. Similarly, the values of $M_{r < r_*}$ above the horizontal line, which are shaded by slanted lines, are ruled out as the horizontal line represents the observationally inferred limit on the mass within the central 10 pc region of the centre of the galaxy. This rules out the combinations of $m$ and ${\hat \lambda}_{ini}$ which correspond to the portion of the curves which lie within either of these regions.
It can be seen that the upper limit on allowed $m$ based on central mass considerations is smaller than the upper limit on $m$ obtained from accretion time considerations. This corresponds to the row with first entry ``0.0" in table~\ref{table:constraints}. 

It is easy to see from figure~\ref{M87WithBH} as well as table~\ref{table:constraints} that when the scalar self-interactions are allowed, these constraints change - the sign and strength of ${\hat \lambda}_{ini}$ determine the amount of this change.
For sufficiently large positive values of ${\hat \lambda}_{ini}$, there may be no lowering of the allowed upper limit on $m$ from that found from accretion considerations. On the other hand, for negative values of ${\hat \lambda}_{ini}$, the lowering of the allowed upper limit on $m$ is more than the lowering for the case considered in \cite{Davies:2019wgi} i.e. the upper limits on $m$ become more stringent.
 
\begin{table}\centering
\begin{tabular}{|cr|c|}
\hline
~~~~~~~${\hat \lambda}_{ini}$ &~~~~~ & excluded $m$ \\ 
\hline
\hline
 &~~~~~& \\
~~~~~-1.0 &~~~~~ & $ m \geq 10^{-22.68} $ eV\\
~~~~~-0.5 &~~~~~& $ m \geq 10^{-22.38} $ eV \\ 
~~~~~0.0 &~~~~~& $m > 10^{-22.18}$ eV\\
~~~~~0.5 &~~~~~ & $m > 10^{-22.06}$ eV\\
~~~~~1.0 &~~~~~ & $m > 10^{-22.06}$ eV\\
\hline
\end{tabular}
\caption{The excluded values of ULDM mass $m$, obtained by the method outlined, get modified in the presence of scalar field DM self-interactions.}
\label{table:constraints}
\end{table}

\subsubsection{Allowed regions in $\lambda - m$ plane}

\begin{figure}
\centering
\begin{subfigure}
\centering
  \includegraphics[width=.85\linewidth]{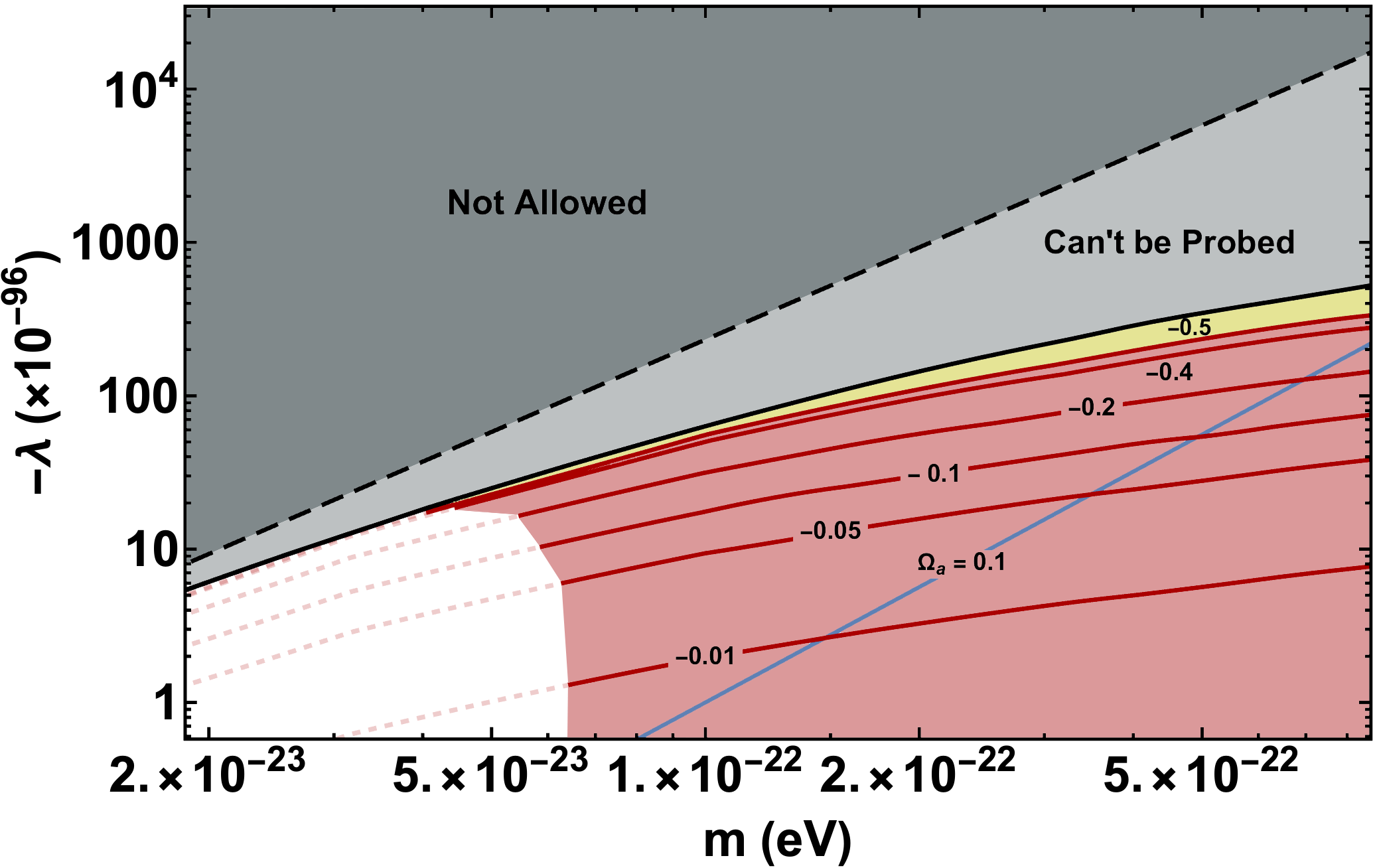}
  \caption{For a given galaxy (such as M87), for every choice of ${\hat \lambda}_{ini}$ of interest, we obtain a curve in 
  $\lambda - m$ plane. In this case, we show the results for negative $\lambda$ i.e. attractive self-interactions. The solid blue line marked ``$\Omega_a = 0.1$" corresponds to the situation in which all of DM consists of Ultra-Light Axions. Note that we have only shown results for a few negative values of ${\hat \lambda}_{ini}$. The dashed line corresponds to eq.~\ref{lambda_constraint}. See main text for discussion about the various regions such as the inaccessible region in this plot.}
  \label{Excluded_-}
\end{subfigure}

\vspace{0.5cm}

\begin{subfigure}
\centering
  \includegraphics[width=.85\linewidth]{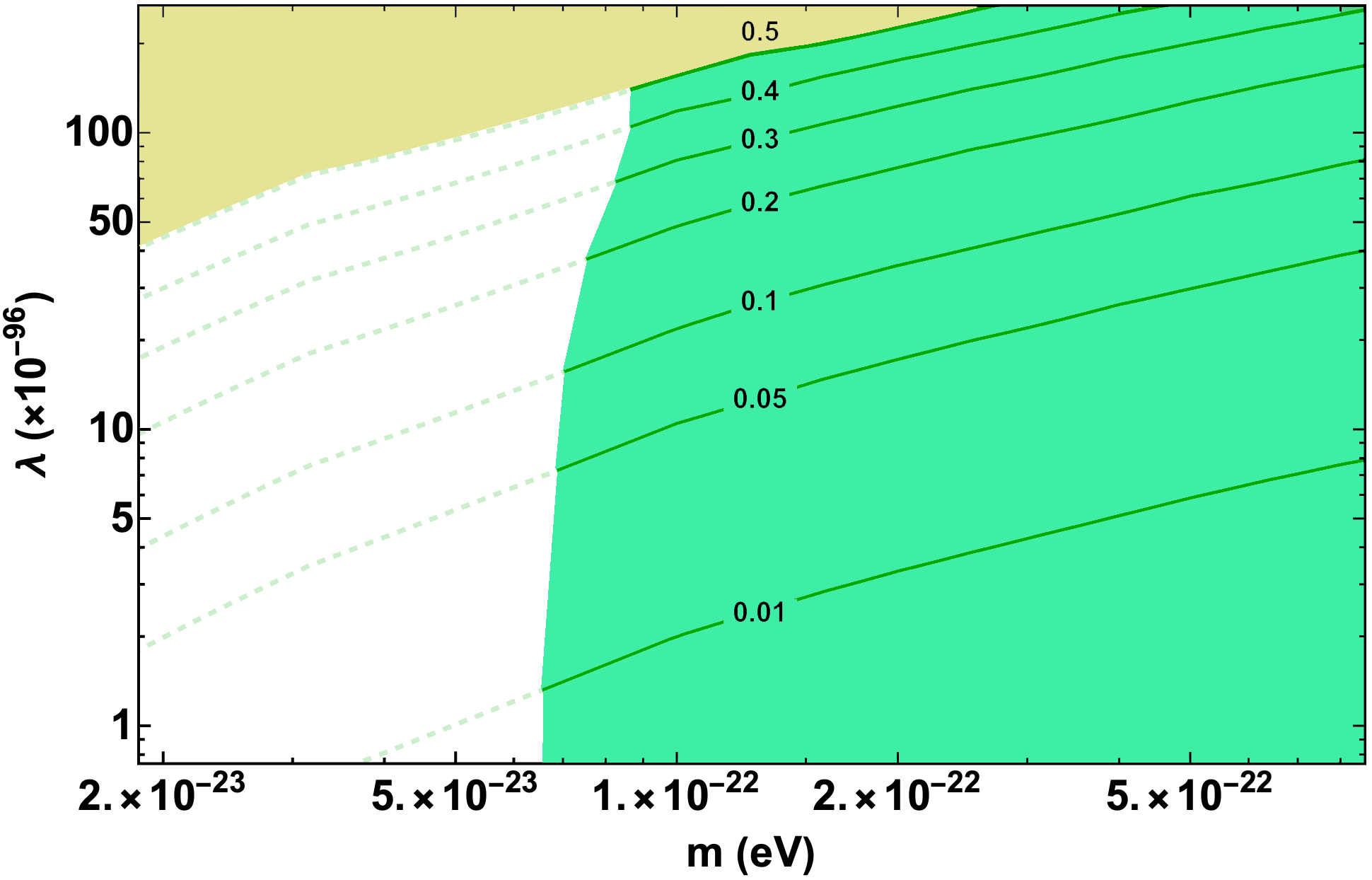}
  \caption{The same as the previous figure but for positive self-interactions. Note that, for the same $|{\hat \lambda}_{ini}|$, the curves in this plot are higher as compared to figure \ref{Excluded_-} and there is no inaccessible region (see the main text for details). }
  \label{Excluded_+}
\end{subfigure}
\label{fig:test}
\end{figure}

We found the impact of ${\cal O}(1)$ values of ${\hat \lambda}_{ini}$ on the mass within central region of the soliton core. 
Using the constraints on $m$ and ${\hat \lambda}_{ini}$ and using eq.~(\ref{eq:lambda}) and eq.~(\ref{eq:lambdac}), we can find the constraints in $\lambda - m$ plane, i.e. we can identify which regions of the $\lambda - m$ parameter space get excluded by the data.
Every curve in figure~\ref{M87WithBH} is for a fixed object (i.e. fixed $M_{\rm halo}$) and for a fixed value of ${\hat \lambda}_{ini}$. Corresponding to each such curve in figure~\ref{M87WithBH}, eq.~(\ref{eq:lambda}) and eq.~(\ref{eq:lambdac}) [or, see eq.~(\ref{eq:lambda0})] suggest that there will be a curve in $\lambda - m$ plane, as is shown in figures~\ref{Excluded_-} and \ref{Excluded_+}. 
Note that figure~\ref{Excluded_-} is obtained for the case of negative (i.e. attractive) scalar self-interactions while figure~\ref{Excluded_+} is obtained for the case of positive (i.e. repulsive) scalar self-interactions. 
Note that the origin of the shapes of the curves in these two  figures is explained in appendix~\ref{sec:m&lambda}.

Since a part of each curve in figure~\ref{M87WithBH} gets excluded, a portion of each of the curves in figures~\ref{Excluded_-} and \ref{Excluded_+} also gets excluded - the excluded portion of each curve is shown as a solid curve while the allowed portion is shown as a dashed curve.
When we consider a whole family of curves corresponding to different values of ${\hat \lambda}_{ini}$, we end up excluding regions in $\lambda - m$ plane.

From figures~\ref{Excluded_-} and \ref{Excluded_+}, we can see that a large region in $\lambda-m$ plane gets excluded by the observational constraints on the mass within 10 pc of the centre of M87 galaxy and accretion time constraints. Note that, for $m \sim 10^{-22}$ eV, the typical values of $\lambda$ which we probe are $\lambda \sim - 10^{-96}$, which correspond to a scattering length of $a_s = \frac{\hbar}{mc}\frac{\lambda}{32\pi} \sim 10^{-82}$ m. 
In eq.~\eqref{eq:lambdac}, we had seen that, for $m \sim10^{-22}$ eV, $\hat \lambda$ of ${\cal O}(1)$ corresponds to $\lambda$ of the order of $10^{-98}$, but, scaling transformations, with scaling parameter $s \sim 100$ (see figure~\ref{s_m} in appendix~\ref{sec:m&lambda}), through eq.~(\ref{eq:lambda}) causes the probed values of $\lambda$ to be two orders of magnitude higher (see also, eq.~(\ref{eq:lambda0})).

As is argued in appendix~\ref{sec:m&lambda}, for a given $|{\hat \lambda}_{ini}|$, the value of the scaling parameter $s$ for positive interactions is always going to be greater than the value of the scaling parameter $s$ for negative interactions. This, along with eq.~(\ref{eq:lambda}) and eq.~(\ref{eq:lambdac}) [or, see eq.~(\ref{eq:lambda0})], explains why, for the same $|{\hat \lambda}_{ini}|$, the curves in figure~\ref{Excluded_+} are higher than the corresponding curves in figure~\ref{Excluded_-}.
Again, as is argued in appendix~\ref{sec:m&lambda}, for negative interactions, for any given object (i.e. DM halo), there will be regions in the $\lambda-m$ plane which can not be probed using the method we have presented. For M87 observations, this region in figure~\ref{Excluded_-} is above the topmost solid black curve and is shaded gray. 

Note that in figure~\ref{Excluded_-}, we only show results assuming that ${\hat \lambda}_{ini}$ is within the range $- \frac{1}{2} \le {\hat \lambda}_{ini} < 0 $ while in figure~\ref{Excluded_+}, we show results assuming that ${\hat \lambda}_{ini}$ is within the range $ \frac{1}{2} \ge {\hat \lambda}_{ini} > 0 $. 
As we saw in section~\ref{sec:typical}, this range corresponds to gravity being stronger than the scalar self-interactions.
If we attempt to obtain similar curves for ${\hat \lambda}_{ini} > \frac{1}{2}$, they will simply lie above the curve labelled ``$0.5$" in figure~\ref{Excluded_+}. This is why the yellow shaded portion above the curve marked ``$0.5$" in figure~\ref{Excluded_+} is the region in which gravity becomes weaker than repulsive scalar self-interactions.
On the other hand, for reasons discussed in appendix~\ref{sec:m&lambda}, for attractive self-interactions, as we attempt to obtain similar curves for ${\hat \lambda}_{ini} < - \frac{1}{2}$, some part of the curves could be below the curve labelled ``$-0.5$" in figure~\ref{Excluded_-}.

\subsubsection{Probing mass and self-coupling of ULAs}

For a canonically normalised field describing an axion or Axion-Like Particle (ALP), \cite{Marsh:2015xka}, the potential has a residual discrete shift symmetry, when all the higher harmonics could be ignored, the potential is of the form
\begin{equation}
 U(\varphi) = m_a^2 f^2 \left[ 1 - \cos \left( \frac{\varphi}{f} \right) \right] \; ,
\end{equation}
where, $f$ is the ``decay constant" of the axion or ALP and using this potential, the quartic self coupling shall be given by
$\lambda_a = - \left( \frac{m_a}{f} \right)^2 $.
For axions with $f \sim {\cal O}(10^{17})~ {\rm GeV}$ and $m_a \sim 10^{-22}~ {\rm eV}$ (hence called Ultra-light axion or ULA) one finds that $\lambda_a \sim - 10^{-96}$, which is a dramatically small negative number.
Furthermore, by calculating the relic abundance of axion DM one finds that, if all of DM consists of axions, the corresponding density is $\Omega_a \sim 0.1\left( \frac{f}{10^{17} {\rm GeV} } \right)^2 \left( \frac{m}{10^{-22} {\rm eV} } \right)^{1/2}$ \cite{Hui:2016ltb}
where, $\Omega_a = \rho_a / \rho_{\rm crit}$, which implies that 
\begin{equation}\label{eq:axion_imp}
\Omega_a \sim 0.1\left(\frac{m}{10^{-22}\ \text{eV}}\right)^{5/2} \left( \frac{10^{-96}}{ -\lambda } \right) \; .
\end{equation}
Thus, in $\log \lambda - \log m $ plane, fixed $\Omega_a$ (such as 0.1) corresponds to a line with slope 5/2. This is shown by the solid blue line labelled ``$\Omega_a = 0.1$" in figure~\ref{Excluded_-}. Needless to say, since the self-coupling of the axion is negative, there will be no axion line in figure~\ref{Excluded_+}.

For sufficiently large strength of attractive self-interactions, gravity and self-interactions can no longer be balanced by ``quantum pressure" and hence there is no stable soliton solution possible \cite{Levkov:2016rkk, Chavanis:2011mrr}. This implies that, for a stable soliton, there is a maximum possible value of soliton mass, denoted by $M_{\text{cr}}$, given by \cite{Chavanis:2011mrr,Levkov:2016rkk}

\begin{equation}\label{critical_mass}
    M_{\text{cr}}
    \approx 10.2 \frac{m_{\text{pl}}}{(-\lambda)^{1/2}} \; .
\end{equation}
For any galaxy, the empirical mass of the soliton is given in the dimensionful form as $M_{\text{emp}} = \frac{\hbar c \hat{M}_{\text{emp}}}{G m}$, where $\hat{M}_{\text{emp}}$ is given by eq.~(\ref{eq:memp}). Rearranging  eq.~(\ref{critical_mass}), one can obtain the maximum $\lambda$ allowed such that the corresponding soliton mass is always less than $M_{\text{cr}}$. For M87, with $M_{\text{halo}} = 2\times 10^{14}\ M_\odot$ we get
\begin{equation}\label{lambda_constraint}
    - \lambda \lesssim 2.34\times 10^{-94}\left(\frac{m}{10^{-22}\ \text{eV}}\right)^2 \; .
\end{equation}
The dashed line in figure 
\ref{Excluded_-} corresponds to the value of $- \lambda$ above which there can be no stable soliton. It is worth noting that the constraints we obtain for attractive self interactions are for allowed range of values of the self-coupling.

Finally, using eq.~(\ref{eq:lambda}) and eq.~(\ref{eq:lambdac}) [or, see eq.~(\ref{eq:lambda0})], one can easily show that, for ${\hat \lambda}_{ini}$ of ${\cal O}(1)$, the scaling parameter $s \sim {\cal O}(10)$. Notice that too small values of $s$, smaller than 10 or so, will take us out of the domain of applicability of our formalism. In particular, when $s$ becomes ${\cal O}(1)$, general relativistic effects can no longer be neglected and we have to include all the higher order terms we ignored in reliably modelling the solitons (see section~\ref{sec:theoretical}). 

But since the scaling parameter does depend on the mass of scalar $m$, ${\hat \lambda}_{ini}$, as well as the galaxy itself (through e.g.~$M_{\rm halo}$), it is expected that there will be some galaxies for which the mass and couplings of an ultra light axion-like scalar field which forms all of DM will be probed. In this context, we would like to point the reader to appendix~\ref{sec:m&lambda} especially to figure~\ref{s_m}. 
For the case of M87 galaxy, the corresponding results are shown in figure~\ref{Excluded_-} and suggest that if all of dark matter is made of ultra light axions, then axions with mass $m \sim {\cal O}(10^{-22})~{\rm eV}$ are ruled out by the observational constraints on the mass within 10 pc of the centre of M87 galaxy and accretion time constraints. Figure~\ref{Excluded_-} also suggests that if ULAs form all of DM, its mass has to be less than
$ \sim 6 \times 10^{-23}$ eV.

\section{Discussion and conclusions}
\label{sec:conclusion}

In the standard model of cosmology, Dark Matter is conjectured to consist of non-relativistic (hence cold) particles, undergoing classical dynamics and with mean free path large compared to the length scales of interest (i.e. collisionless). 
There exist observations at sufficiently small length scales, e.g. galactic scales, which could be interpreted to imply that this picture of dark matter may be inaccurate or even invalid. Given the fact that this is inferred from observations at galactic scales, it is sensible to find ways to test models of DM at those scales. In particular, its worth asking whether observations of the myriads of galaxies can teach us something about the fundamental parameters in the Lagrangian of DM.

With this context in mind, we began by indicating how the Gross-Pitaevskii-Poisson equations follow from the weak field, slow variation and non-relativistic limit of a classical field theory with a canonical self-interacting scalar field and Einstein gravity. We then assumed that the astrophysical DM is in fact the self-interacting classical scalar field in this theory. We then modelled the cores of DM halos by solving the Gross-Pitaevskii-Poisson equations in section~\ref{sec:machinery} and looked for stationary, spherically symmetric, nodeless solutions. 

This was done by first finding the solutions with an additional normalisation condition ${\hat \phi}(0) \sim 1$. It was then argued that such solutions were both physically unrealistic and untrustworthy (which we called ``theoretical solitons"). We then used the scaling symmetry in the Gross-Pitaevskii-Poisson equations to obtain realistic, trustworthy solutions from the theoretical solitons. In section~\ref{sec:scaling} and section~\ref{sec:parameter}, we described the strategy used to obtain the amount of scaling which is required to arrive at a model of the DM halo core for M87 galaxy.

In addition, the DM halo of interest may have a central black hole (BH) present - we model it as a point particle in Newtonian gravity. 
In the absence of a central black hole or scalar self-interactions, the density profile of the scalar field has a universal form \cite{Schive:2014dra}. In \cite{Davies:2019wgi}, it was shown that the presence of a central supermassive black hole squeezes the soliton. 
We found that the presence of scalar self-interactions leads to many interesting possibilities; e.g. the presence of self interactions could squeeze the soliton if the interactions are attractive ($\lambda < 0$) and stretch it if the self interactions are repulsive ($\lambda > 0$). This can be easily seen in figure~\ref{fig:densityprofile}.

Needless to say, if the self-interaction strength is too low, it will have no observable effects, but, if it is too high, the structure of DM halos will be governed not by gravity but by self-interactions alone - this is a scenario which we have not explored in this work. When we impose this additional requirement that scalar self-interactions do not overtake gravity - this implies that the parameter $|{\hat \lambda}_{ini}|$ should be less than $1/2$ (see section~\ref{sec:typical}), and we get a curve in the parameter plane such that all constraints of interest are below this curve (the curve marked $-0.5$ in figure~\ref{Excluded_-} and the curve marked $+0.5$ in figure~\ref{Excluded_+}).
As we argued in appendix \ref{sec:m&lambda}, for a given value of $|{\hat \lambda}_{ini}|$, the probed $\lambda$ obtained from eq.~(\ref{eq:lambda0}) for attractive self-interactions will be more than the probed $\lambda$ for negative interactions. We also found that, for every halo, if the scalar self-interactions are attractive, there will be regions in $\lambda-m$ plane, which can never be probed by the method presented in this paper. 

Using the dependence of accretion time of scalar field into super massive black holes on $m$ and using the upper limits on the amount of DM concentrated near the central black holes, \cite{Davies:2019wgi} placed constraints on the mass of Ultra Light DM particles. In the presence of self-interactions of the scalar field, an analysis, presented in section~\ref{sec:M&D}, which was a modification of the one in \cite{Davies:2019wgi}, implies that the lower limit on the mass $m$ depends upon the sign and strength of the self-interactions (see figure~\ref{M87WithBH} as well as table~\ref{table:constraints}). 
This consideration is important while comparing the limits on $m$ obtained by the methods of \cite{Davies:2019wgi} with the limits on $m$ obtained elsewhere in the literature.

Before closing, it is important to note a few caveats and sources of errors and uncertainties. 
This method is obviously currently limited by various uncertainties associated with (a) theoretical modelling of accretion of scalar field, (b) observational limits on the masses of SMBH and the amount of DM contained within the central region etc. Furthermore, as we noted in section~\ref{sec:scaling} the empirical relations used to obtain the amount of scaling need not be applicable for many galaxies. We have used them to illustrate the ideas we present here. As discussed in the beginning of section~\ref{sec:constraints}, the accretion time considerations we used are based on the results of accretion modelling for a scalar field without self-interactions. Similarly, as we noted in section~\ref{sec:M&D}, we have not been committed to any specific definition of the exact size of theoretical soliton which we took to be ${\cal O}(m^{-1})$ i.e. there is an ${\cal O}(1)$ uncertainty in the size of theoretical soliton. Throughout this paper, we illustrated all our ideas using the numerical estimates for M87 galaxy as discussed in section~\ref{sec:m87}. These numbers (e.g. the mass of soliton) have uncertainties associated with them and this can cause the actual numbers in our plots to change considerably.

Some of the problems due to these uncertainties are expected to be ameliorated in the near future e.g. the theoretical modelling of the accretion process, the observational limits on the DM mass within the central region of galaxies, the observational limits on the mass of super-massive black holes etc are all expected to improve. 
Needless to say, the region of the $\lambda-m$ plane we constrain has been explored by other approaches \cite{Fan:2016rda,Cembranos:2018ulm,Suarez:2016eez,Chavanis:2020rdo,Desjacques:2017fmf,Urena-Lopez:2019xri,Delgado:2022vnt}. Our primary focus here has been developing the method and finding the constraints using an illustrative set of parameter values. With that in mind, we 
can conclude that (a) for every galaxy, there will be a region in the $\lambda - m$ parameter plane which will get excluded by accretion time considerations and estimates of mass within central regions of a galaxy; (b) by including the effects of self-interactions of the ultra light scalar DM under consideration, the values of self-coupling which can be probed is extremely small i.e. ${\cal O}(10^{-96})$ - i.e. of the order of those corresponding to ultra-light axions with Planckian decay constants;
(c) the probed values of self-coupling for positive self-interactions are higher than those of negative self-interactions,
(d) our analysis suggests that if axions form all of DM, then, axionic DM with $m \sim {\cal O}(10^{-22)}~{\rm eV}$ is ruled out by central mass estimates for M87 galaxy - its mass has to be less than $ \sim 6 \times 10^{-23}$ eV (e) for negative self-interactions, for any given DM halo, there will be regions in $\lambda - m$ plane which can never be probed using the methods described in this paper.

\acknowledgments
The authors would like to thank Devarshee Sandilya for initial collaboration on this project. The authors would also like to thank Professor Raghavan Rangarajan and Professor Alexander Dolgov for discussions at various stages of the work. This work is supported by Department of Science and Technology, Government of India under Indo-Russian call for Joint Proposals (DST/INT/RUS/RSF/P-21). Bihag Dave acknowledges support from the above mentioned project as a Junior Research Fellow. The work of Koushik Dutta is partially supported by the grant MTR/2019/000395 and  Core Research Grant CRG/2020/004347 funded by SERB, DST, Government of India. 
Finally, the authors would also like to thank the anonymous referee for insightful comments and suggestions.

\appendix

\section{Details of scaling}

\subsection{Empirical correlations}
\label{sec:scaling}
For any given value of mass of halo $M_{\rm halo}$, there exist empirical relations such as the soliton mass $-$ halo mass relation (see \cite{Schive:2014hza} and the discussion in section 3.1 of \cite{Davies:2019wgi}), which can be used to obtain the observed value of dimensionless soliton mass ${\hat M}_{\rm emp}$ from the halo mass $M_{\rm halo}$

 \begin{equation} \label{eq:memp}
 \hat{M}_\text{emp} = 5.45\times 10^{-3} \left(\frac{M_\text{halo}}{2\times 10^{14}\ M_\odot}\right)^{1/3} \; .
\end{equation}

It is important to note that, as pointed out in ref.~\cite{Bar:2019pnz}, the soliton mass - halo mass relation was obtained from simulations \cite{Schive:2014dra} in the absence of any SMBHs. Furthermore, the mass range of halos for which the simulations are carried out is $M_\text{halo} \sim (10^9 - 5\times 10^{11})\ M_\odot$ for $m \approx 10^{-22}\ \text{eV}$. Thus, for a galaxy with SMBH at the centre such that $M_\text{halo} \gtrsim 10^{12}\ M_\odot$ and for dark matter mass $m$ values much different from $10^{-22}\ \text{eV}$, strictly speaking, this relation is not applicable. Despite this, we proceed with the use of this relation in the rest of the work since our main focus is only on illustrating the method we present. 

Similarly, there exists an empirical relation which relates halo mass $M_{\rm halo}$ to the mass of the central BH $M_{\bullet}$ (see \cite{Bandara:2009bhm} and the discussion in section 3.2 of \cite{Davies:2019wgi}). Using eq.~(\ref{eq:alpha}), one can find the corresponding observed value of the parameter $\hat \alpha$ i.e. ${\hat \alpha}_{\rm emp} (m, M_{\rm halo})$:

 \begin{equation} \label{eq:alphaemp}
\hat{\alpha}_\text{emp} = 1.18\times 10^{-2}\left(\frac{M_\text{halo}}{2\times 10^{14} \ M_\odot}\right)^{1.55}\left(\frac{m}{10^{-22}\ \text{eV}}\right) \; .
\end{equation}

In summary, given the halo mass $M_{\rm halo}$ one can use such correlations to determine the observed dimensionless soliton mass ${\hat M}_{\rm emp}$ using eq.~(\ref{eq:memp}). Similarly, given the halo mass $M_{\rm halo}$ and the ULDM mass $m$ of interest, one can determine the observed value of the quantity ${\hat \alpha}$ from eq.~(\ref{eq:alphaemp}).

In some cases e.g. for the Milky Way galaxy or M87 galaxy, both $M_\bullet$ and $M_{\rm halo}$  are well measured and hence one can use this measured value of $M_\bullet$ (along with $m$ of interest) to determine ${\hat \alpha}_{\rm emp}$ from eq.~(\ref{eq:alpha}).

\subsection{Determination of the scaling parameter ``$s$"}
\label{sec:parameter}
 
One can use the following strategy to find the amount of scaling: 

\begin{itemize}
 \item If a particular object such as a galactic halo is to be used to constrain the parameters of ULDM, its own mass, $M_{\rm halo}$, is already fixed. In order to proceed, let us also fix the mass of ULDM particle i.e. $m$.
 \item Start with some initial value of ${\hat \lambda}$ called ${\hat \lambda}_{ini}$ (note that this value will correspond to some value of self-coupling $\lambda$ as defined by eq.~(\ref{eq:lambdac}) in terms of $m$).
 \item Determine ${\hat M}_{\rm emp}$ using eq.~(\ref{eq:memp}).
 Similarly, determine ${\hat \alpha}_{\rm emp}$ from either eq.~(\ref{eq:alphaemp}), or, if the mass of central black hole is already known, directly using eq.~(\ref{eq:alpha}).
 \item Choose an initial value for $\hat \alpha$, call this ${\hat \alpha}_{ini}$ and solve eqs.~(\ref{eq:GP_dimless}) and (\ref{eq:P_dimless}) to find the corresponding ${\hat M}$ and call it ${\hat M}_{ini}$.
 \item Find the value of scaling parameter such that the mass of the soliton after scaling is equal to the empirical mass of the soliton (using eq.~(\ref{eq:Mtrans})) i.e. find $s$ such that
\begin{equation}
s = \frac{{\hat M}_{ini}}{{\hat M}_{\rm emp}} \; .
\end{equation}
 \item This scaling should also transform $\hat \alpha$, 
 so, transform $\hat \alpha$ using this scaling parameter, 
 i.e. find ${\hat \alpha}_{trans} = s^{-1} {\hat \alpha}_{ini}$. 
We then ask ourselves: is ${\hat \alpha}_{trans} = {\hat \alpha}_{\rm emp}$? 
In practice, we have to ask whether they are equal to some desired accuracy i.e. we have to ask whether the following is true:
\begin{equation}
 \left[ \frac{ {\hat \alpha}_{trans} - {\hat \alpha}_{\rm emp} }{ {\hat \alpha}_{\rm emp} } \right] < 0.05 \; ,
\end{equation}
(assuming $5 \%$ accuracy).
 \item If the answer is `yes', we have found the correct ${\hat \alpha}_{ini}$ and the corresponding $s$ is the desired scaling parameter. On the other hand, if it is `no', we try another value of ${\hat \alpha}_{ini}$. 
 \item Once the desired ``$s$" is obtained, the value of self-coupling probed using the object of mass $M_{\rm halo}$, and for ULDM of mass $m$, is give by (eq.~(\ref{eq:lambdatrans}))  
\begin{equation}\label{eq:lambda}
 {\hat \lambda} = s^2 {\hat \lambda}_{ini} \; ,
\end{equation}
 which can be used to determine the corresponding value of probed $\lambda$ using eq.~(\ref{eq:lambdac}).
\end{itemize}
By following this procedure, given $M_{\rm halo}$, $m$ and ${\hat \lambda}_{ini}$, we can find the initial $\hat \alpha$ i.e. ${\hat \alpha}_{ini}$ which we must start with, such that, after scaling by an amount $s$, the dimensionless soliton mass is 
${\hat M}_{\rm emp}$ and after scaling, the $\hat \alpha$ parameter takes up its desired empirical value ${\hat \alpha}_{\rm emp}$.
Thus, $m$, ${\hat \lambda}_{ini}$ and $M_{\rm halo}$ are the final free variables.

This explains how the various curves in figure~\ref{fig:densityprofile} have been obtained. The halo mass is fixed to $M_{\rm halo} = 2 \times 10^{14} M_{\odot}$ while the ULDM mass is $m = 10^{-22} ~ {\rm eV}$. Different curves correspond to various values of ${\hat \lambda}_{ini}$ chosen which imply different values of the self coupling $\lambda$. Note that each curve corresponds to a different value of the scaling parameter ``$s$". 
 In particular, as one can see from figure~\ref{s_m} in appendix~\ref{sec:m&lambda}, the value of $s$ for ${\hat \lambda}_{ini} = +1$ is $s_+ = 181.5$, while the value of $s$ for ${\hat \lambda}_{ini} = -1$ is $s_- = 66.6$. The square of the ratio of these numbers gives the ratio of $\lambda$ corresponding to ${\hat \lambda}_{ini} = +1$ and ${\hat \lambda}_{ini} = -1$ (which is mentioned in the caption of figure~\ref{fig:densityprofile}). The reasoning in appendix~\ref{sec:m&lambda} explains why one expects $s_+$ to be always larger than $s_-$.

\section{The shapes of the curves in figure~\ref{M87WithBH}}
\label{sec:Shape1}

In this section, we try to understand why the shapes of the various curves in figure \ref{M87WithBH} are what they are.

Let us first think about the case of a theoretical soliton without self-interactions and without a central black hole. It is worth recalling that the size of theoretical soliton is $R \sim {\cal O}(m^{-1})$ and its mass is $M \sim {\cal O}(M_{pl}^2\cdot  m^{-1})$. Thus, as we increase $m$, the soliton gets smaller and lighter. Let us start with DM of extremely small mass, the corresponding soliton shall be very big (and very heavy) and only a very small fraction of the soliton will be inside a sphere of radius $r_*$ of the centre of the soliton. Within this small central region of the soliton, if we assume that its density is constant, $\rho_0$, the mass within $r_*$ will be $M_{r<r_*} \sim \rho_0 r_*^3$. Now, since $\rho_0 \sim M R^{-3}$, the density itself goes as $m^{2}$, so, $M_{r<r_*} \sim m^2 r_*^3$. This is the reason why the left part of a plot of $\ln M_{r<r_*}$ against $\ln m$ will appear to be a straight line with slope 2. As we increase $m$, the soliton keeps getting smaller, at some point, the entire soliton is inside $r_*$, so, $M_{r<r_*}$ becomes equal to $M$ which we know goes as $m^{-1}$ - this explains why the right part of a plot of $\ln M_{r<r_*}$ against $\ln m$ will be a straight line with slope -1. 

When we need to go from theoretical soliton to real soliton, we need to use the scaling parameter. It turns out that in the absence of a central black hole and in the absence of scalar self-interactions, the scaling parameter $s$ itself does not depend on $m$. Thus, the argument in the last paragraph is applicable to real solitons also. In the presence of central black hole or self-interactions (or both), $s$ will be $m$ dependent and strictly speaking, the arguments of the last paragraph will not apply.
The presence of central black hole squeezes the soliton. In addition, the presence of attractive or repulsive self-interactions can further compress or expand it. Finally, the density of the soliton is not exactly the same at all radii. These additional effects slightly modify the shapes of the curves in figure \ref{M87WithBH} but the basic explanation remains the same.

Now let us understand the distinction between the curve corresponding to ${\hat \lambda} = 0$, the solid curve labelled ``0" in figure \ref{M87WithBH}, and the various curves corresponding to the presence of interactions, ${\hat \lambda} \neq 0$. 
In the rightmost part of the figure~\ref{M87WithBH}, irrespective of the sign of the interactions or the presence of a central black hole, the behaviour is identical to the behaviour in ${\hat \lambda} = 0$ case. This is because, as we saw above, the rightmost part of figure~\ref{M87WithBH} corresponds to large $m$ i.e. large ${\hat r}_*$ for which the condition ${\hat r}_* < s {\hat r}_{\rm sol} $ is satisfied and the entire soliton is inside the region of interest. Thus, the integral in eq.~(\ref{eq:M_full}) will be the same for all the cases as the upper limit of the integral can be replaced by $\infty$.

To understand the left part of figure~\ref{M87WithBH} in the presence of interactions, first recall that ${\hat \lambda} > 0$ corresponds to repulsive self interactions while ${\hat \lambda} < 0$ corresponds to attractive self-interactions.
For ${\hat \lambda} > 0$, the soliton is bigger and hence for every value of $m$ to the left of the peak, less mass will be enclosed within the region of interest compared to the case of no interactions. This causes the corresponding curve to be lower than the curve for ${\hat \lambda} = 0$ case. 
On the other hand, for ${\hat \lambda} < 0$ or in the presence of a central black hole, the soliton will be smaller and hence for every value of $m$ to the left of the peak, more mass will be enclosed within the region of interest compared to the case of no interactions and this will cause the corresponding curve to be higher than the curve for ${\hat \lambda} = 0$ case. 

It is useful to understand in a different way how this works out. For the unscaled (i.e. theoretical soliton), the integrand ${\hat \phi}^2 {\hat r}^2$ in eq.~(\ref{eq:mhat}) can be plotted against ${\hat r}$.  One finds that the curve corresponding to ${\hat \lambda} > 0$ lies above the curve corresponding to ${\hat \lambda} < 0$. The dimensionless soliton mass ${\hat M}$ will be the area under this curve. Thus, we expect ${\hat M}_{ini}^+ > {\hat M}_{ini}^-$, i.e. the dimensionless soliton mass for unscaled soliton for the case of repulsive self-interactions is greater than that in the case of attractive self-interactions. This particular point can also be understood from figure~\ref{Mhatinivsalphahatini} and the discussion in section~\ref{sec:m&lambda} below.
Consequently, eq.~(\ref{eq:s}) implies that $s_+ > s_-$ i.e. the scaling parameter for ${\hat \lambda} > 0$ case is more than that in ${\hat \lambda} < 0$ case. This will cause the integrand in eq.~(\ref{eq:M_limited}) for the two cases to scale as $1 / s_+^2$ and $1 / s_-^2$ i.e. the integrand for ${\hat \lambda} > 0$ case will decrease more than the integrand for the ${\hat \lambda} < 0$ case. This causes the curves for repulsive interactions in figure~\ref{M87WithBH} to be below the curves for attractive interactions.

\section{Probed $\lambda$ for a chosen $m$ and for a given object}
\label{sec:m&lambda}

It is very useful to understand the shapes of the curves and excluded regions in $\lambda-m$ plane using the formalism we have developed. Similarly, one might wonder, for a given object (fixed $M_{\rm halo}$) and a chosen range of $m$ values, what typical values of $\lambda$ one can probe or test using the method presented here. In particular, one might wonder whether there are any regions in the $\lambda-m$ plane which can not be probed using the present method.
In this section, we clarify all of these issues.

\begin{figure}
  \includegraphics[width = 0.9\textwidth]{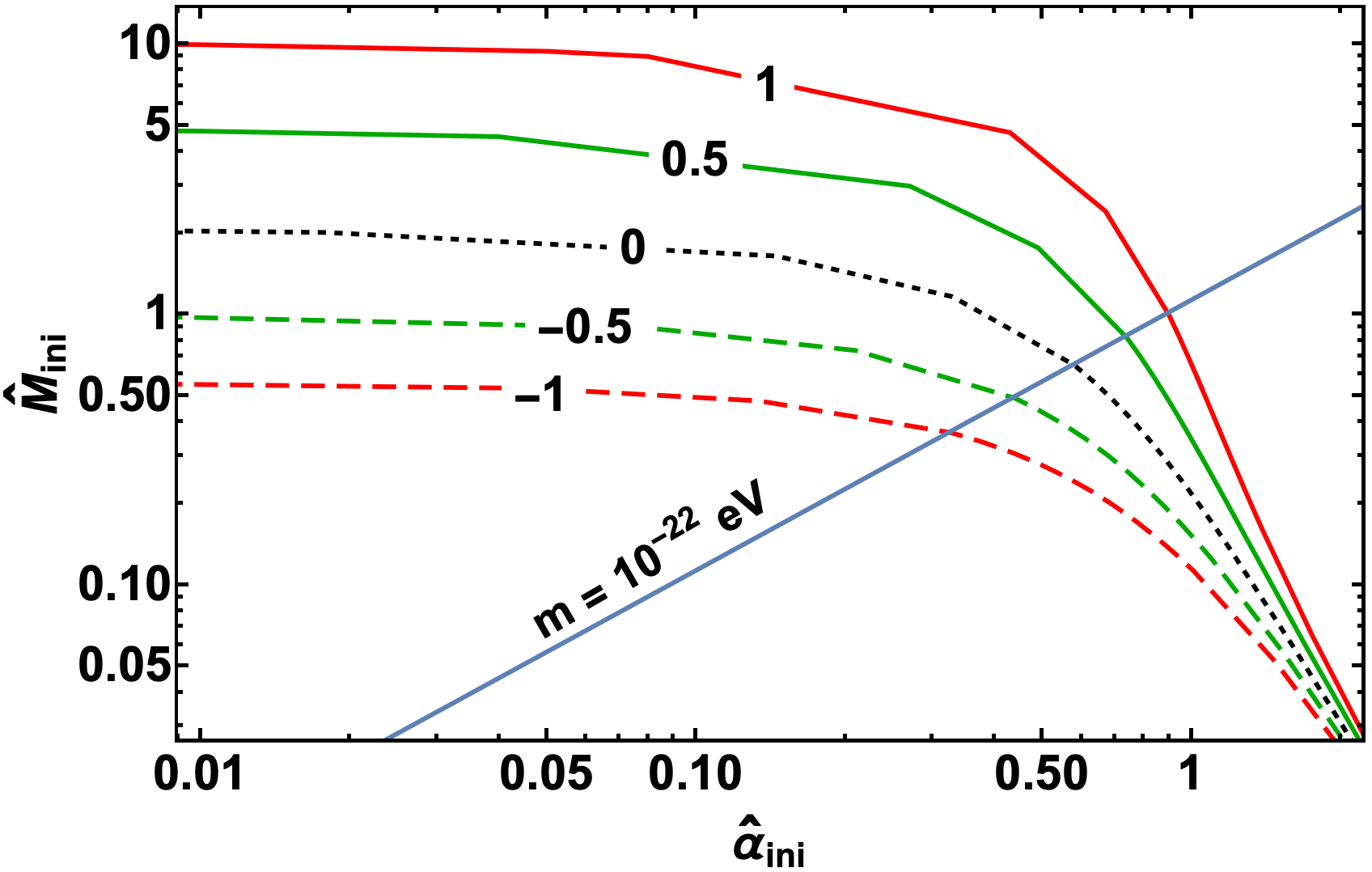}
  \caption{The dependence of $\ln {\hat M}_{ini}$ on $\ln {\hat \alpha}_{ini}$ for various choices of ${\hat \lambda}_{ini}$ for M87 galaxy. From this figure, it is easy to see that ${\hat M}_{ini}^+ > {\hat M}_{ini}^-$. Given this, eq.~(\ref{eq:s}) implies that $s_+ > s_-$, thus, for a given value of $|{\hat \lambda}_{ini}|$, the probed $\lambda$ obtained from eq.~(\ref{eq:lambda0}) for attractive self-interactions will be more than the probed $\lambda$ for negative interactions. The straight line is the one obtained from the right hand side of eq \ref{eq:LHS=RHS}, its slope is 1 and intercept is given by eq (\ref{eq:obs}).
}
  \label{Mhatinivsalphahatini}
\end{figure}

Once the values of $\hat \alpha$ and ${\hat \lambda}$ in eqs.~(\ref{eq:GP_dimless}) and (\ref{eq:P_dimless}) are known, let us say ${\hat \alpha} = {\hat \alpha}_{ini}$ and ${\hat \lambda} = {\hat \lambda}_{ini}$, we can solve for ${\hat \phi}({\hat r})$ and find the dimensionless mass of the soliton i.e. ${\hat M} = {\hat M}_{ini}$. In figure~\ref{Mhatinivsalphahatini}, we show the dependence of ${\hat M}_{ini}$ on ${\hat \alpha}_{ini}$ for a few possible values of ${\hat \lambda}_{ini}$. This plot will help us understand many issues in this section. 

From eq.~(\ref{eq:lambdac}) and eq.~(\ref{eq:lambda}), we can conclude that, for a given object (i.e. fixed $M_{\rm halo}$), the probed self coupling value $\lambda$ depends on $m$, ${\hat \lambda}_{ini}$ and the value of scaling parameter $s$ since

\begin{equation}\label{eq:lambda0}
 \lambda = 8 s^2 {\hat \lambda}_{ini} \left( \frac{m}{M_{pl}} \right)^2 \; .
\end{equation}
If $s$ were a constant, for a given ${\hat \lambda}_{ini}$ (and a given object i.e. given $M_{\rm halo}$), the plot of $\ln \lambda$ against $\ln m$ shall be a straight line with slope 2. In reality, as we shall now argue, the scaling parameter $s$ itself depends on all the three independent variables $M_{\rm halo}$, $m$ and ${\hat \lambda}_{ini}$.

We saw in section~\ref{sec:parameter} that, in order to turn an unrealistic soliton into a realistic model of the core of DM halo, we must satisfy the requirement, eq.~(\ref{eq:s}),
\begin{equation}\label{eq:LHS=RHS}
 {\hat M}_{ini} ({\hat \alpha}_{ini} , {\hat \lambda}_{ini}) = \left( \frac{ {\hat M}_{\rm emp} }{{\hat \alpha}_{\rm emp}} \right) {\hat \alpha}_{ini} \; ,
\end{equation}
where, because of eqs.~(\ref{eq:memp}) and (\ref{eq:alphaemp}), the ratio in the brackets on the RHS goes as $M_{\rm halo}^{-1.22} \cdot m^{-1}$.
The LHS of this equation is what is shown as the family of curves in figure \ref{Mhatinivsalphahatini}. 
The right hand side of this equation suggests that, in figure~\ref{Mhatinivsalphahatini}, for a given value of $m$, the observationally consistent values of $\ln {\hat M}_{ini}$, as a function of $\ln {\hat \alpha}_{ini}$, should be straight lines with slope 1 and intercept which is roughly given by (using eqs.~(\ref{eq:memp}) and (\ref{eq:alphaemp})),

\begin{equation}\label{eq:obs}
\text{Intercept} = \ln\left[0.47\left(\frac{M_\text{halo}}{2\times 10^{14} \ M_\odot}\right)^{-1.22}\left(\frac{m}{10^{-22}\ \text{eV}}\right)^{-1}\right]
\end{equation}
For M87, for DM of mass $m = 10^{-22}~{\rm eV}$, the corresponding line is shown in figure~\ref{Mhatinivsalphahatini}.
Thus, for a chosen value of ${\hat \lambda}_{ini}$, the correct value of ${\hat M} = {\hat M}_{ini}$ and ${\hat \alpha}_{ini}$ will be the one at the intersection of the curves in figure~\ref{Mhatinivsalphahatini} and straight lines with slope 1 and intercept given by eq.~(\ref{eq:obs}).

\subsection{The shapes of curves in figures~\ref{Excluded_-} and \ref{Excluded_+}} 

\begin{figure}
  \includegraphics[width = 0.95\textwidth]{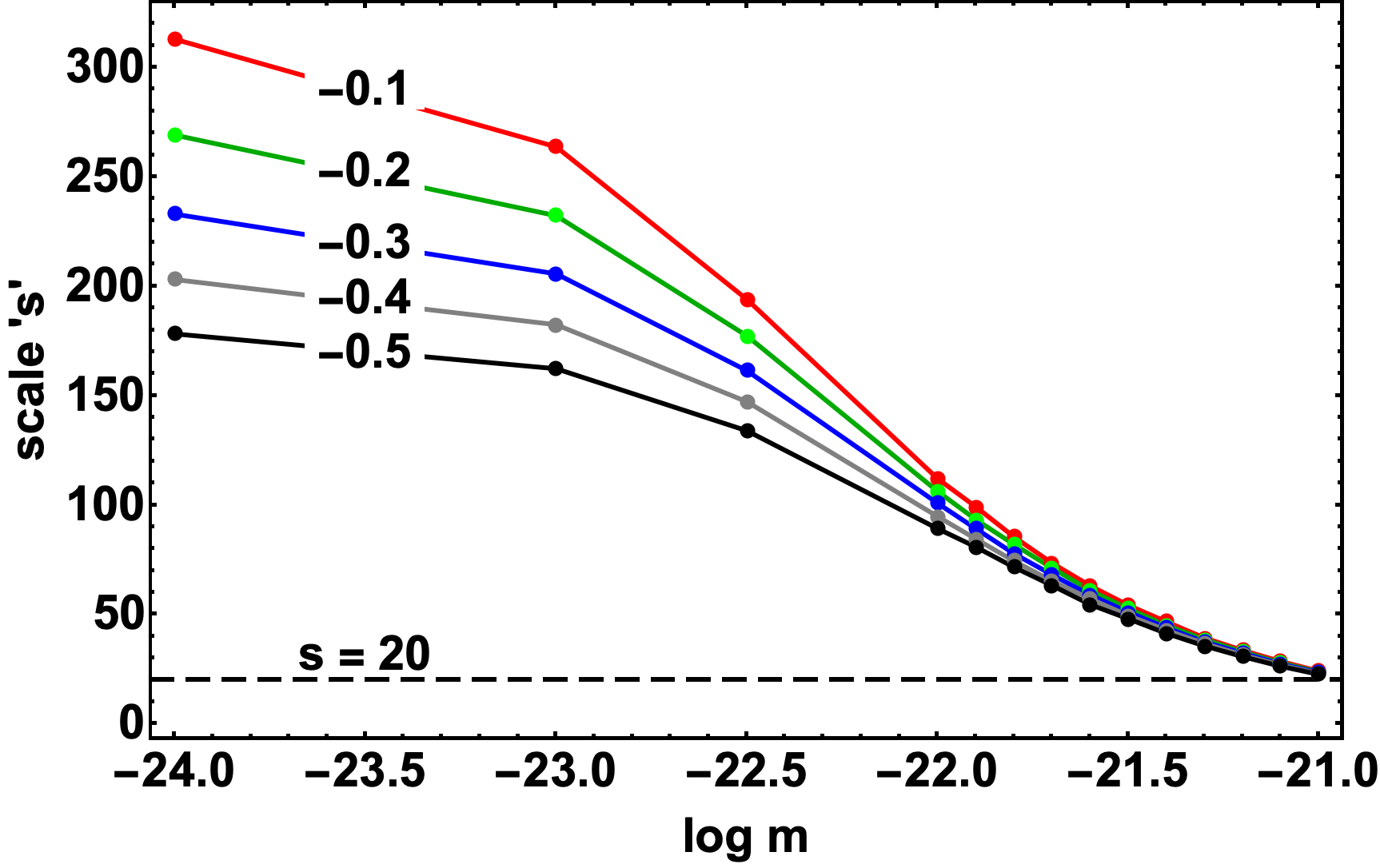}
  \caption{The plot of scaling parameter $s$ against $m$ for various values of ${\hat \lambda}_{ini}$ for M87 galaxy. As long $s$ is sufficiently large compared to 1, the formalism in this paper is trustworthy.
}
  \label{s_m}
\end{figure}

As we increase $m$, eq~(\ref{eq:obs}) informs us that the intercept of the line representing the RHS of eq.~(\ref{eq:LHS=RHS}) decreases, this means that in figure~\ref{Mhatinivsalphahatini} we end up exploring the rightmost regions of the plot - the corresponding ${\hat \alpha}_{ini}$ is larger and ${\hat M}_{ini}$ is smaller. Since $M_{\rm halo}$ is fixed, ${\hat M}_{\rm emp}$ remains the same, eq.~(\ref{eq:s}) then implies that $s$ must decrease. This can be verified from the curves in figure~\ref{s_m}.

For very small values of $m$, the intercept of the line (using eq.~(\ref{eq:obs})) becomes too large and the line intersects the curves in figure~\ref{Mhatinivsalphahatini} in the leftmost, flatter part. Under this circumstance, if we increase $m$ slightly, the intercept decreases and the line shifts only slightly downwards, causing the intersection point of the line and the curve to be shifted only slightly to the right. Since we are in the leftmost, flatter part of the curves in figure~\ref{Mhatinivsalphahatini}, this means that the corresponding value of ${\hat M}_{ini}$ doesn't decrease appreciably. Thus, from eq.~(\ref{eq:s}), we find that $s$ shall not change much - something easy to verify from figure~\ref{s_m}. 

Thus, the quantity $\partial s / \partial m$ has less negative values for smaller values of $m$ and more negative values for larger values of $m$. Now, using eq.~(\ref{eq:lambda0}), one can easily show that
\begin{equation}
 \frac{\partial \lambda}{\partial m} = 16 {\hat \lambda}_{ini} \left[  \left( \frac{m s^2 }{M_{pl}^2} \right) + s  \left( \frac{\partial s}{\partial m} \right) \left( \frac{m}{M_{pl}} \right)^2  \right] \; ,
\end{equation}
i.e.
\begin{equation}
 \frac{\partial \ln | \lambda| }{\partial \ln m} = 2 \left[  1 +  \frac{\partial \ln s}{\partial \ln m} \right] \; .
\end{equation}
Thus, the plot of $\ln | \lambda| $ against $\ln m$ shall be a straight line with slope 2 for smaller values of $m$ while the slope will decrease from 2 for larger values of $m$. This explains the basic shape of the curves in figures~\ref{Excluded_-} and \ref{Excluded_+}.

\subsection{Change in probed $\lambda$ as ${\hat \lambda}_{ini}$ changes (for fixed $M_{\rm halo}$ and $m$)}

Let us first focus our attention on the case with ${\hat \lambda}_{ini} < 0$ i.e. attractive self-interactions.

\begin{figure}
  \includegraphics[width = 0.95\textwidth]{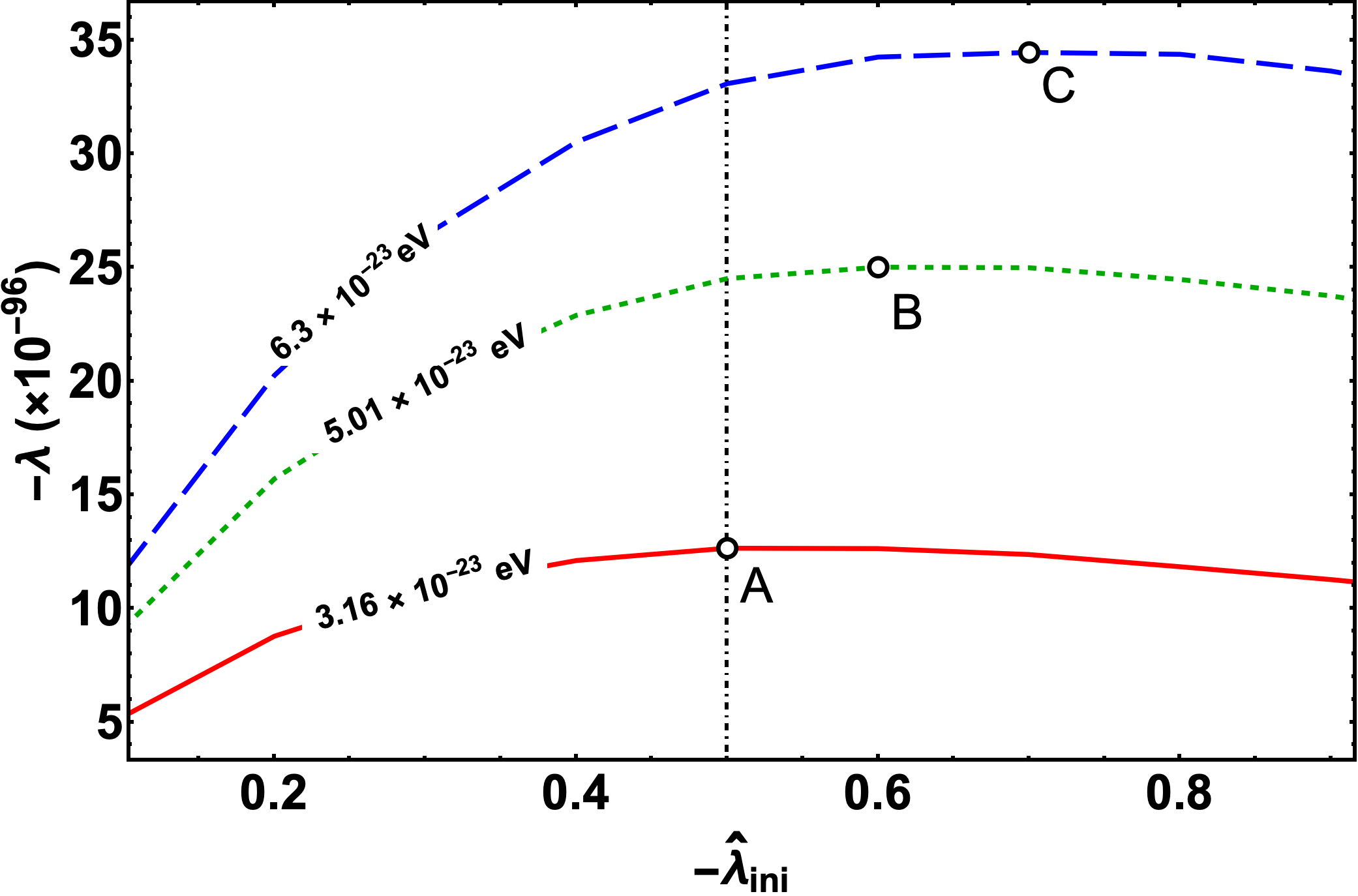}
  \caption{The plot of $\lambda$ against $\hat{\lambda}_{ini}$ for three different values of $m$ as marked on each curve. Points A, B and C correspond to the local maximum for each curve. The vertical dot-dashed line corresponds to $\hat{\lambda}_{ini} = -0.5$.}
  \label{lambda0-lambdac}
\end{figure}

When ${\hat \lambda}_{ini}$ becomes more negative, from figure~\ref{Mhatinivsalphahatini}, this leads to a lower value of ${\hat M}_{ini}$, thus, eq.~(\ref{eq:s}) then implies that $s$ must decrease - this is true for all values of $m$ but the decrease in $s$ is more for smaller $m$ and less for larger $m$. 
Thus, for negative self-interactions, in eq (\ref{eq:lambda0}), the function $s^2$, which depends on ${\hat \lambda}_{ini}$ falls very quickly at smaller $m$ and less quickly at larger $m$.
Since eq.~(\ref{eq:lambda0}) involves a product of a quickly decreasing function $s^2 ({\hat \lambda}_{ini})$ and an increasing function $|{\hat \lambda}_{ini}|$, this implies that, as we increase $| {\hat \lambda}_{ini} |$, the probed values of $\ln | \lambda| $ will first increase and then decrease. This means that, for a given $m$, curves corresponding to ever increasing values of $| {\hat \lambda}_{ini} |$ will have a $\ln | \lambda| $ value which will first increase and then decrease, this can be seen from figure~\ref{lambda0-lambdac}.  

Thus, for a given object (i.e. fixed $M_{halo}$), $\lambda$ depends on $m$ and ${\hat \lambda}_{ini}$. The dependence of $\lambda$ on ${\hat \lambda}_{ini}$ for different values of $m$ is shown in figure~\ref{lambda0-lambdac} and is completely expected given the above explanation.
What we have argued is that for any given object (such as M87 galaxy), for every choice of $m$, there is always a local maximum in the plot of probed $\lambda$ against ${\hat \lambda}_{ini}$. If the value of $\lambda$ corresponding to the local maximum is called $\lambda_{\rm max}$, we can find such a value for every $m$. If, for a given object with a fixed $M_{\rm halo}$, one plots $\lambda_{\rm max}$ against $m$, in such a plot, for a given $m$, any value of $\lambda > \lambda_{\rm max}$ for that $m$ can not be probed for any choice of ${\hat \lambda}_{ini}$ by the method we have presented in this paper. This is the origin of the inaccessible region in $\lambda-m$ plane for attractive self-interactions. 

From figure~\ref{Mhatinivsalphahatini} and eq.~(\ref{eq:lambda0}), it is easy to see that this will not happen for positive self-interactions.

\subsection{Change in probed $\lambda$ as $M_{\rm halo}$ increases (for fixed ${\hat \lambda}_{ini}$  and $m$)}

As halo mass is increased, from eq.~(\ref{eq:obs}), the intercept of the line of interest will decrease, causing ${\hat M}_{ini}$ to decrease and ${\hat M}_{\rm emp}$ to increase. Thus, $s$ must decrease too quickly causing the probed value of $\lambda$ to decrease too. Thus, one can probe smaller values of self-coupling using heavier halos provided the corresponding value of $s$ doesn't become too small for the formalism to be applicable.

In summary, $s$ decreases for increasing values of $M_{\rm halo}$, $m$ and $| {\hat \lambda}_{ini} |$ for negative self-interactions.  If $s$ becomes too small, the formalism we are working with, in which the GR effects have been ignored, will not be valid. This imposes another practical restriction on the region of $\ln \lambda - \ln m$ plane that we can probe using our methods.
For the purpose of illustrating the basic idea, we restrict our attention to the range of parameter values for which the scaling parameter always stays bigger than 20, as is shown in figure~\ref{s_m}.

\end{document}